\documentclass[11pt,italian,a4widepaper]{article}

\usepackage{graphics,graphicx}
\usepackage{amssymb}
\usepackage{amsmath}
\usepackage{graphicx}
\usepackage{enumerate}
\usepackage[dvips]{color}

\begin{document}

\newcommand{\singlespace}{\baselineskip=12pt
\lineskiplimit=0pt \lineskip=0pt }
\def\ds{\displaystyle}

\newcommand{\beq}{\begin{equation}}
\newcommand{\eeq}{\end{equation}}
\newcommand{\lb}{\label}
\newcommand{\beqar}{\begin{eqnarray}}
\newcommand{\eeqar}{\end{eqnarray}}
\newcommand{\und}{\underline}
\newcommand{\diam}{\stackrel{\scriptscriptstyle \diamond}}

\newcommand{\Ehat}{\hat{E}}
\newcommand{\Ahat}{\hat{A}}
\newcommand{\khat}{\hat{k}}
\newcommand{\muhat}{\hat{\mu}}
\newcommand{\mc}{M^{\scriptscriptstyle C}}
\newcommand{\mt}{M^{\scriptscriptstyle T}}
\newcommand{\mei}{M^{\scriptscriptstyle M,EI}}
\newcommand{\mec}{M^{\scriptscriptstyle M,EC}}
\newcommand{\mw}{M^{\scriptscriptstyle W}}

\newenvironment{sistema}%
{\left\lbrace\begin{array}{@{}l@{}}}%
{\end{array}\right.}

\def\ob{{\, \underline{\otimes} \,}}
\def\scalp{\mbox{\boldmath$\, \cdot \,$}}
\def\gdp{\makebox{\raisebox{-.215ex}{$\Box$}\hspace{-.778em}$\times$}}
\def\bob{\makebox{\raisebox{-.215ex}{$\Box$}\hspace{-.73em}$\scalp$}}

\def\c{{\circ}}

\def\bA{\mbox{\boldmath${\it A}$}}
\def\ba{\mbox{\boldmath${\it a}$}}
\def\bB{\mbox{\boldmath${\it B}$}}
\def\bb{\mbox{\boldmath${\it b}$}}
\def\bC{\mbox{\boldmath${\it C}$}}
\def\bc{\mbox{\boldmath${\it c}$}}
\def\bD{\mbox{\boldmath${\it D}$}}
\def\bd{\mbox{\boldmath${\it d}$}}
\def\bE{\mbox{\boldmath${\it E}$}}
\def\be{\mbox{\boldmath${\it e}$}}
\def\bF{\mbox{\boldmath${\it F}$}}
\def\bff{\mbox{\boldmath${\it f}$}}
\def\bG{\mbox{\boldmath${\it G}$}}
\def\bg{\mbox{\boldmath${\it g}$}}
\def\bH{\mbox{\boldmath${\it H}$}}
\def\bh{\mbox{\boldmath${\it h}$}}
\def\bi{\mbox{\boldmath${\it i}$}}
\def\bj{\mbox{\boldmath${\it j}$}}
\def\bK{\mbox{\boldmath${\it K}$}}
\def\bk{\mbox{\boldmath${\it k}$}}
\def\bL{\mbox{\boldmath${\it L}$}}
\def\bl{\mbox{\boldmath${\it l}$}}
\def\bM{\mbox{\boldmath${\it M}$}}
\def\bm{\mbox{\boldmath${\it m}$}}
\def\bN{\mbox{\boldmath${\it N}$}}
\def\bn{\mbox{\boldmath${\it n}$}}
\def\b0{\mbox{\boldmath${0}$}}
\def\bo{\mbox{\boldmath${\it o}$}}
\def\bP{\mbox{\boldmath${\it P}$}}
\def\bp{\mbox{\boldmath${\it p}$}}
\def\bQ{\mbox{\boldmath${\it Q}$}}
\def\bq{\mbox{\boldmath${\it q}$}}
\def\br{\mbox{\boldmath${\it r}$}}
\def\bR{\mbox{\boldmath${\it R}$}}
\def\bS{\mbox{\boldmath${\it S}$}}
\def\bs{\mbox{\boldmath${\it s}$}}
\def\bT{\mbox{\boldmath${\it T}$}}
\def\bt{\mbox{\boldmath${\it t}$}}
\def\bU{\mbox{\boldmath${\it U}$}}
\def\bu{\mbox{\boldmath${\it u}$}}
\def\bv{\mbox{\boldmath${\it v}$}}
\def\bV{\mbox{\boldmath${\it V}$}}
\def\bw{\mbox{\boldmath${\it w}$}}
\def\bW{\mbox{\boldmath${\it W}$}}
\def\by{\mbox{\boldmath${\it y}$}}
\def\bX{\mbox{\boldmath${\it X}$}}
\def\bx{\mbox{\boldmath${\it x}$}}

\def\bbD{\overline{\bD}}
\def\bbL{\overline{\bL}}
\def\bbW{\overline{\bW}}

\def\bbeta{\mbox{\boldmath${\beta}$}}
\def\bepsilon{\mbox{\boldmath${\epsilon}$}}
\def\bvarepsilon{\mbox{\boldmath${\varepsilon}$}}
\def\bsigma{\mbox{\boldmath${\sigma}$}}
\def\bphi{\mbox{\boldmath${w}$}}
\def\bzeta{\mbox{\boldmath${\zeta}$}}

\def\Id{\mbox{\boldmath${\it I}$}}
\def\balpha{\mbox{\boldmath${\alpha}$}}
\def\bbeta{\mbox{\boldmath${\beta}$}}
\def\bGamma{\mbox{\boldmath${\Gamma}$}}
\def\bDelta{\mbox{\boldmath${\Delta}$}}
\def\bkappa{\mbox{\boldmath $\kappa$}}
\def\btau{\mbox{\boldmath $\tau$}}
\def\bnu{\mbox{\boldmath $\nu$}}
\def\bchi{\mbox{\boldmath${\chi}$}}
\def\bxi{\mbox{\boldmath${ \xi}$}}
\def\bXi{\mbox{\boldmath${\it  \Xi}$}}
\def\bsigma{\mbox{\boldmath${\sigma}$}}
\def\bSigma{\mbox{\boldmath${\Sigma}$}}
\def\bupsilon{\mbox{\boldmath $\upsilon$}}
\def\bgamma{\mbox{\boldmath $\gamma$}}
\def\bTheta{\mbox{\boldmath $\Theta$}}

\def\tr{{\sf tr}}
\def\dev{{\sf dev}}
\def\div{{\sf div}}
\def\Div{{\sf Div}}
\def\Grad{{\sf Grad}}
\def\grad{{\sf grad}}
\def\Lin{{\sf Lin}}
\def\Orth{{\sf Orth}}
\def\Unim{{\sf Unim}}
\def\Sym{{\sf Sym}}

\def\msm{\mbox{${\mathsf m}$}}

\def\msM{\mbox{${\mathsf M}$}}
\def\msS{\mbox{${\mathsf S}$}}

\def\forA{\mathbb A}
\def\forB{\mathbb B}
\def\forC{\mathbb C}
\def\forE{\mathbb E}
\def\forL{\mathbb L}
\def\forN{\mathbb N}
\def\forR{\mathbb R}

\def\capA{\mbox{\boldmath${\mathsf A}$}}
\def\capB{\mbox{\boldmath${\mathsf B}$}}
\def\capC{\mbox{\boldmath${\mathsf C}$}}
\def\capD{\mbox{\boldmath${\mathsf D}$}}
\def\capE{\mbox{\boldmath${\mathsf E}$}}
\def\capF{\mbox{\boldmath${\mathsf F}$}}
\def\capG{\mbox{\boldmath${\mathsf G}$}}
\def\capH{\mbox{\boldmath${\mathsf H}$}}
\def\capI{\mbox{\boldmath${\mathsf I}$}}
\def\capK{\mbox{\boldmath${\mathsf K}$}}
\def\capL{\mbox{\boldmath${\mathsf L}$}}
\def\capM{\mbox{\boldmath${\mathsf M}$}}
\def\capR{\mbox{\boldmath${\mathsf R}$}}
\def\capW{\mbox{\boldmath${\mathsf W}$}}

\def\C{\mbox{\boldmath${\mathcal C}$}}
\def\E{\mbox{\boldmath${\mathcal E}$}}

\def\mA{\mbox{${\mathcal A}$}}
\def\mB{\mbox{${\mathcal B}$}}
\def\mC{\mbox{${\mathcal C}$}}
\def\mD{\mbox{${\mathcal D}$}}
\def\mE{\mbox{${\mathcal E}$}}
\def\mF{\mbox{${\mathcal F}$}}
\def\mG{\mbox{${\mathcal G}$}}
\def\mH{\mbox{${\mathcal H}$}}
\def\mI{\mbox{${\mathcal I}$}}
\def\mJ{\mbox{${\mathcal J}$}}
\def\mK{\mbox{${\mathcal K}$}}
\def\mL{\mbox{${\mathcal L}$}}
\def\mM{\mbox{${\mathcal M}$}}
\def\mQ{\mbox{${\mathcal Q}$}}
\def\mR{\mbox{${\mathcal R}$}}
\def\mS{\mbox{${\mathcal S}$}}
\def\mT{\mbox{${\mathcal T}$}}
\def\mV{\mbox{${\mathcal V}$}}
\def\mY{\mbox{${\mathcal Y}$}}
\def\mZ{\mbox{${\mathcal Z}$}}

\def\AAM{{\it Adv. Appl. Mech. }}
\def\AMM{{\it Acta Metall. Mater. }}
\def\ARMA{{\it Arch. Rat. Mech. Analysis }}
\def\AMR{{\it Appl. Mech. Rev. }}
\def\CMAME {{\it Comput. Meth. Appl. Mech. Engrg. }}
\def\CMT {{\it Cont. Mech. and Therm.}}
\def\CRAS{{\it C. R. Acad. Sci., Paris }}
\def\EFM{{\it Eng. Fract. Mech. }}
\def\EJMA{{\it Eur.~J.~Mech.-A/Solids }}
\def\IMA{{\it IMA J. Appl. Math. }}
\def\IJES{{\it Int. J. Engng. Sci. }}
\def\IJMS{{\it Int. J. Mech. Sci. }}
\def\IJNME{{\it Int. J. Numer. Meth. Eng. }}
\def\IJNAMG{{\it Int. J. Numer. Anal. Meth. Geomech. }}
\def\IJP{{\it Int. J. Plasticity }}
\def\IJSS{{\it Int. J. Solids Struct. }}
\def\IngA{{\it {Ing. Archiv }}}
\def\JACS{{\it J. Am. Ceram. Soc. }}
\def\JAM{{\it J. Appl. Mech. }}
\def\JAP{{\it J. Appl. Phys. }}
\def\JE{{\it J. Elasticity }}
\def\JM{{\it J. de M\'ecanique }}
\def\JMPS{{\it J. Mech. Phys. Solids. }}
\def\MOM{{\it Mech. Materials }}
\def\MRC{{\it Mech. Res. Comm. }}
\def\MSE{{\it Mater. Sci. Eng. }}
\def\MMS{{\it Math. Mech. Solids }}
\def\MPCPS{{\it Math. Proc. Camb. Phil. Soc. }}
\def\PRSA{{\it Proc. R. Soc. Lond., Ser. A}}
\def\PRSL{{\it Proc. R. Soc. Lond. }}
\def\QAM{{\it Quart. Appl. Math. }}
\def\QJMAM{{\it Quart. J. Mech. Appl. Math. }}
\def\ZAMP{{\it Z. angew. Math. Phys. }}
\def\ZAMM{{\it Z. angew. Math. Mech. }}


\def\salto#1#2{%
[\mbox{\hspace{-#1em}}[#2]\mbox{\hspace{-#1em}}]}


\title{\bf Mindlin second-gradient elastic properties \\ from dilute two-phase Cauchy-elastic composites\\
Part I: Closed form expression for the effective higher-order constitutive tensor}
\author{M. Bacca, D. Bigoni\footnote{Corresponding author}, F. Dal Corso \& D. Veber \\
Department of Civil, Environmental and Mechanical Engineering\\
University of Trento, \\ via Mesiano 77, I-38123 Trento, Italy\\
e-mail: mattia.bacca@ing.unitn.it, bigoni@unitn.it,\\
francesco.dalcorso@unitn.it, daniele.veber@unitn.it}
\date{}
\maketitle

\begin{abstract}

It is shown that second-order homogenization of a Cauchy-elastic dilute suspension of randomly distributed inclusions
yields an equivalent second gradient (Mindlin) elastic material.
This result is valid for both plane and three-dimensional problems and
extends earlier findings by Bigoni and Drugan (Analytical derivation of Cosserat moduli
via homogenization of heterogeneous elastic materials. \emph{J. Appl. Mech.}, 2007, \textbf{74}, 741--753) from several points of view: (i.)
the result holds for anisotropic phases with
 spherical or circular ellipsoid of inertia;
(ii.) the displacement boundary conditions considered in the homogenization procedure is independent of the characteristics
of the material; (iii.) {\it a perfect energy match is found} between heterogeneous and equivalent materials
(instead of an optimal bound).
The constitutive higher-order tensor defining the equivalent Mindlin solid is given
in a surprisingly simple formula. Applications, treatment of material symmetries and positive definiteness of the effective higher-order constitutive tensor are deferred to Part II of the present article.

\end{abstract}


\section{Introduction}

Due to the lack of a characteristic length, local constitutive models are
unsuitable for mechanical applications at the micro- and nano-scale,
since size-effects evidenced by experiments cannot be described and the
modelling fails when large strain gradient are present, as in the case of
shear band formation (Dal Corso and Willis, 2011). Therefore, many nonlocal
models have been formulated and
developed, starting from the pioneering work by the Cosserat brothers (1909)
and by Koiter (1964) and Mindlin (1964).
Despite their evident connection to the microstructure, nonlocal models
are usually introduced in a phenomenological way, so that attempts of
explicitly relating the microstructure to nonlocal effects are scarce
(theoretical considerations were developed by Achenbach and Hermann, 1968; Beran and McCoy, 1970; Boutin, 1996; Dal Corso and Deseri, 2013;
Forest and Trinh, 2011; Li, 2011; Pideri and
Seppecher, 1997; Wang and Stronge, 1999;
numerical approaches were given by Auffray et al. 2010; Forest, 1998;
Ostoja-Starzewski et al. 1999; Bouyge et al. 2001; experiments were provided by Anderson and Lakes, 1994;
Buechner and Lakes, 2003; Lakes, 1986; Gauthier, 1982).

Bigoni and Drugan (2007) have provided a technique to identify Cosserat
constants from homogenization of a heterogeneous Cauchy elastic solid.
Their approach shows how a nonlocal material can be realized starting from
a \lq usual' Cauchy elastic composite and opens the way to the
practical realization of nonlocal materials. Their methodology has two
important limitations, namely, that (i.) the obtained characteristic
lengths for the Cosserat material {\it do not} allow a complete match of
the elastic energies between the Cauchy heterogeneous and the Cosserat
homogeneous materials, but {\it minimize} the energy difference between these two, and (ii.) that
the homogenization is performed by imposing boundary displacements
depending on the Poisson's ratio of the material (so that the boundary conditions
considered are not exactly equal). These two limitations are
overcome in the present article, by using a higher-order \lq
Mindlin' nonlocal elastic material which provides a perfect match between
the elastic energies of
a dilute suspension of Cauchy-elastic inclusions (randomly distributed in a Cauchy-elastic matrix)
and a homogeneous non-local elastic material, obtained through application of
the same displacement field at the boundary. Moreover, although our results
remain confined to the dilute assumption, we also generalize Bigoni and
Drugan (2007) by relaxing (iii.) the restriction of isotropy and (iv.) the
shape of the inclusions, which may now have a generic form (though subject
to certain geometrical restrictions to be detailed later).

Description of the proposed identification procedure of the Mindlin elastic constants and the relative closed-form formulae
 are reported in this article, while a discussion about positive-definiteness, material symmetries and
applications to explicit cases are deferred to Part II.

\section{Preliminaries on Second-Gradient Elasticity (SGE)}

The equations are briefly introduced governing the equilibrium of the second-gradient elastic (SGE) solid proposed by Mindlin (1968)
that will be employed in the homogenization procedure.

Considering a quasi-static deformation process, defined by the displacement field $\bu$ (function of the position $\bx$),
 the primary kinematical quantities of the SGE are defined as
\beq\lb{kinematical}
\varepsilon_{ij}=\frac{u_{i,j}+u_{j,i}}{2}, \qquad
\chi_{ijk}=u_{k,ij},
\eeq
where a comma denotes differentiation, the indices range between 1 and $N$
(equal to 2 or 3, depending on the space dimensions of the problem considered),
and $\bvarepsilon$ and $\bchi$ are the (second-order) strain and the (third-order) curvature tensor fields,
respectively, satisfying the following symmetry properties
\beq
\varepsilon_{ij}=\varepsilon_{ji}, \qquad \chi_{ijk}=\chi_{jik}.
\eeq
Defining the statical entities Cauchy stress $\sigma_{ij}$=$\sigma_{ji}$ and double stress $\tau_{ijk}$=$\tau_{jik}$,
respectively work-conjugate to the kinematical entities
$\bvarepsilon$ and $\bchi$, eqn (\ref{kinematical}), the principle of virtual work can be written for a solid occupying a domain $\Omega$,
with boundary $\partial \Omega$ and set of edges $\Gamma$, in the absence of body-force as
\beq\label{pvw}
\int_\Omega (\sigma_{ij} \delta\varepsilon_{ij}+ \tau_{ijk}\delta\chi_{ijk})=
\int_{\partial\Omega} (t_i \delta u_i+ T_i D\delta u_i)
+\int_{\Gamma}\Theta_i \delta u_i,
\eeq
where repeated indices are summed, $\bt$ represents
the surface traction (work-conjugate to $\bu$), while $\bT$ and $\bTheta$ denote the generalized
tractions on the surface
$\partial \Omega$ and along the set of edges $\Gamma$ (work-conjugate respectively to $D\bu$ and $\bu$),
and $D=n_l \partial_l$ represents the derivative along
the outward normal direction to the boundary, $\bn$ (definite only on $\partial\Omega$ but not on $\Gamma$).
Through integration by parts, the equilibrium conditions, holding for points within the body $\Omega$, can be obtained as
\beq\lb{indequilibrium}
\partial_j\left(\sigma_{jk}-\partial_i\tau_{ijk}\right) = 0,\\[3 mm]
\qquad \mbox{in } \Omega,
\eeq
while for points on the boundary $\partial \Omega_p$ and along the set of edges $\Gamma_p$,
(where statical conditions are prescribed in terms of $\bt$, $\bT$ and $\bTheta$) as
\beq
\lb{boundaryconditions}
\left\{
\begin{split}
&n_j\sigma_{jk}-n_i n_j D\tau_{ijk}-2 n_j D_i \tau_{ijk}+\left(n_i n_j D_l n_l-D_j n_i\right)\tau_{ijk}= t_k,\\[3 mm]
&n_i n_j \tau_{ijk}=T_{k},
\end{split}
\right. \qquad \mbox{on } \partial\Omega_p,
\eeq
and
\beq
\salto{0.5}{\, e_{mlj} n_i s_m n_l \tau_{ijk} \,}=\Theta_k,
\qquad ~~~ \mbox{on}\, \Gamma_p,
\eeq
where $e_{mlj}$ is the Ricci \lq permutation' tensor, $D_j=\left(\delta_{jl}-n_j n_l\right)\partial_l$,
 $\bs$ is the unit vector tangent to $\Gamma$ and $\salto{0.1}{\cdot}$ represents the jump
of the enclosed quantity, computed with the normals $\bn$ defined on the surfaces intersecting at the edge $\Gamma$.
Finally, kinematical conditions\footnote{In the proposed homogenization procedure only kinematical boundary conditions will be imposed
($\partial \Omega_p\equiv \emptyset$, so that $\partial \Omega_u\equiv\partial \Omega$). } are prescribed for
points on the remaining boundary $\partial \Omega_u\equiv\partial \Omega\backslash\partial \Omega_p$
 as
\beq
\left\{
\begin{array}{lll}
u_i=\overline{u}_i, \\[4mm]
D u_i=\overline{Du}_i,
\end{array}
\right.
\qquad \mbox{on } \partial\Omega_u.
\eeq

Introducing the strain energy density $w^{SGE}=w^{SGE}(\bvarepsilon,\bchi)$, the $\bsigma$ and $\btau$ fields
can be obtained as
\beq\label{potentialpartial}
\sigma_{ij}=\frac{\partial w^{SGE}}{\partial \varepsilon_{ij}},\qquad
\tau_{ijk}=\frac{\partial w^{SGE}}{\partial \chi_{ijk}},
\eeq
so that,  restricting attention to centrosymmetric materials within a linear theory\footnote{
Centrosymmetry is coherent with the fact that the elastic energies at first- and at second- order
are decoupled under the geometrical assumptions that will be introduced in Section 3.1.
}, it follows that
\beq
\lb{defenergysgm}
w^{SGE}(\bvarepsilon,\bchi)=\underbrace{\frac{1}{2}\capC_{ijhk}\varepsilon_{ij}\varepsilon_{hk}}_{w^{SGE,L}(\bvarepsilon)}
+\underbrace{\frac{1}{2}\capA_{ijklmn}\chi_{ijk}\chi_{lmn}}_{w^{SGE,NL}(\bchi)},
\eeq
where $\capC$ and $\capA$ are the local (fourth-order) and non-local
(sixth-order) constitutive tensors, each generating respectively a strain energy
density contribution, say \lq local', $w^{SGE,L}$ (corresponding to the energy stored in a Cauchy material,
$w^{SGE,L}=w^C$) and \lq non-local', $w^{SGE,NL}$.
Therefore, the linear constitutive equations for the
stress and double stress quantities are obtained as
\beq \lb{defconstsgm}
\sigma_{ij}=\capC_{ijhk}\varepsilon_{hk}, \qquad
\tau_{ijk}=\capA_{ijklmn}\chi_{lmn},
\eeq
which, from eqns (\ref{kinematical}) and (\ref{potentialpartial}), have the following symmetries
\beq
\capC_{ijhk}=\capC_{jihk}=\capC_{ijkh}=\capC_{hkij},\qquad
\capA_{ijklmn}=\capA_{jiklmn}=\capA_{ijkmln}=\capA_{lmnijk}.
\eeq
In the case of isotropic response, the constitutive elastic tensors $\capC$ and $\capA$ can be written in the following form
\beq
\lb{isotropy_tensor}
\begin{array}{rll}
\ds\capC_{ijhk}=&\ds\lambda \delta_{ij} \delta_{hk}+\mu (\delta_{ih}\delta_{jk}+\delta_{ik}\delta_{jh}) ,
\\[5mm]
\ds\capA_{ijhlmn}
=&\ds\frac{a_1}{2}
\left[\delta_{ij}\left(\delta_{hl}\delta_{mn}+\delta_{hm}\delta_{ln}\right)
    +\delta_{lm}\left(\delta_{in}\delta_{jh}+\delta_{ih}\delta_{jn}\right)\right]
\\[3mm]
&+\ds\frac{a_2}{2}
\left[\delta_{ih}\left(\delta_{jl}\delta_{mn}+\delta_{jm}\delta_{ln}\right)
    +\delta_{jh}\left(\delta_{il}\delta_{mn}+\delta_{im}\delta_{ln}\right)\right]
\\[3mm]
&+\ds 2\,a_3\left(\delta_{ij}\delta_{hn}\delta_{lm}\right)+a_4
\left(\delta_{il}\delta_{jm}+
     \delta_{im}\delta_{jl}\right)\delta_{hn}
\\[3mm]
&+\ds\frac{a_5}{2}
\left[\delta_{in}\left(\delta_{jl}\delta_{hm}+\delta_{jm}\delta_{hl}\right)
    +\delta_{jn}\left(\delta_{il}\delta_{hm}+\delta_{im}\delta_{hl}\right)\right] ,
\end{array}
\eeq
where $\delta_{ij}$ is the Kronecker delta,
$\lambda$ and $\mu$ are the usual Lam\'{e} constants, defining the local isotropic behavior, while $a_i$ ($i=1,...,5$) are the five
material constants (with the dimension of a force) defining the nonlocal isotropic behavior.
Considering the constitutive isotropic tensors (\ref{isotropy_tensor}), the strain energy density (\ref{defenergysgm}) becomes
\beq
\lb{isoenergysgm}
w^{SGE}(\bvarepsilon,\bchi)=\underbrace{\frac{\lambda}{2} \varepsilon_{ii}\varepsilon_{jj}+
\mu \varepsilon_{ij}\varepsilon_{ij}}_{w^{SGE,L}(\bvarepsilon)}+
\underbrace{\sum_{k=1}^5 a_k \mI_k(\bchi)}_{w^{SGE,NL}(\bchi)},
\eeq
where the invariants $\mI_k(\bchi)$ are
\beq\lb{Inv}\begin{split}
\mI_1(\bchi)&=\chi_{iik}\,\chi_{jkj}(=\chi_{iik}\,\chi_{kjj}) ,\\
\mI_2(\bchi)&=\chi_{iki}\,\chi_{jkj}(=\chi_{kii}\,\chi_{jkj}=\chi_{kii}\,\chi_{kjj}=\chi_{iki}\,\chi_{kjj}),\\
\mI_3(\bchi)&=\chi_{iik}\,\chi_{jjk} , \\
\mI_4(\bchi)&=\chi_{ijk}\,\chi_{ijk}(=\chi_{jik}\,\chi_{ijk}=\chi_{jik}\,\chi_{jik}=\chi_{ijk}\,\chi_{jik}), \\
\mI_5(\bchi)&=\chi_{ijk}\,\chi_{kji}(=\chi_{jik}\,\chi_{kji}=\chi_{jik}\,\chi_{jki}=\chi_{ijk}\,\chi_{kji}),
\end{split}\eeq

so that the linear constitutive relations (\ref{defconstsgm}) reduce  to
\beq \lb{isoconstsgm}\begin{array}{rll}
\sigma_{ij} =&\ds\lambda \varepsilon_{ll} \delta_{ij} + 2\mu \varepsilon_{ij},\\[5mm]
\tau_{ijk} =&\ds\frac{a_1}{2}\left(\chi_{lli}\delta_{jk}+2\chi_{kll}\delta_{ij}+\chi_{llj}\delta_{ik}\right)+
a_2 \left(\chi_{ill}\delta_{jk}+\chi_{jll}\delta_{ik}\right)+2 a_3 \chi_{llk}\delta_{ij}\\[5 mm]
&\ds +2 a_4 \chi_{ijk}+a_5 \left(\chi_{kji}+\chi_{kij}\right).
\end{array}\eeq

Since the invariants defined by eqns (\ref{Inv}) satisfy the following inequalities
\beq\lb{orchite}
\begin{array}{ccc}
2\mI_1(\bchi) + \mI_2(\bchi)+\mI_3(\bchi)\geq 0,\qquad
\mI_2(\bchi) \geq 0, \qquad \mI_3(\bchi) \geq 0,\\[3mm]
\mI_4(\bchi) \geq 0,\qquad \mI_4(\bchi) + \mI_5(\bchi) \geq 0,
\end{array}
\eeq
the positive definiteness condition for the isotropic strain energy density
$w^{SGE}(\bvarepsilon,\bchi)$, eqn (\ref{isoenergysgm}),
corresponds to the usual restraints for the local parameters
(given by the positive definiteness of $w^{SGE,L}(\bvarepsilon)$)
\beq\lb{posdefiso}
3\lambda +2\mu > 0, ~~~ \mu >0,~~~
\eeq
which are complemented by the following conditions (Mindlin and Eshel, 1968) on the nonlocal constitutive parameters
(given by the positive definiteness of $w^{SGE,NL}(\bchi)$)
\beq\lb{posdefiso2}
-a_4<a_5<2 a_4,~~~e_1>0,~~~e_2>0,~~~5e_3^2<2e_1e_2,
\eeq
where
\beq\begin{array}{c}
e_1=-4a_1+2a_2+8a_3+6a_4-3a_5,~~~e_2=5(a_1+a_2+a_3)+3(a_4+a_5),\\[2mm]
e_3=a_1-2a_2+4a_3.
\end{array}\eeq

\section{Homogenization procedure}\lb{homogenizationproceduresect}

The proposed homogenization procedure follows Bigoni and Drugan (2007). In particular, the same\footnote{
Bigoni and Drugan (2007) impose a linear and quadratic displacement field on the boundaries
of the RVE and of the homogeneous equivalent material, whose quadratic part depends on the Poisson's ratio of the material
to which the displacement
is applied, so that the applied displacements are not exactly equal.
Furthermore, the equivalent material considered by Bigoni and Drugan is a non-local Koiter
material (1964), which does not permit the annihilation, but only a minimization
of the elastic energy mismatch between the RVE and the equivalent material.}
 (linear and quadratic) displacement is applied on the boundary of both the representative volume element RVE
and the homogeneous equivalent SGE material.
Then, the equivalent local $\capC^{eq}$ and non-local $\capA^{eq}$ tensors are obtained imposing the vanishing of the
elastic energy mismatch between the two materials.
Since the strain energy in the homogeneous SGE material is given only by the local contribution
when linear displacement boundary condition are applied (because no strain gradient arises),
the equivalent local tensor $\capC^{eq}$ corresponds to that obtained with usual homogenization procedures.
Thus, the remaining unknown of the equivalent SGE material (namely, the non-local equivalent constitutive tensor $\capA^{eq}$) can be obtained
by imposing the vanishing mismatch in strain energy when (linear and) quadratic displacement are considered.
A chief result in the current procedure is that a perfect match in the elastic energies is achieved,
while Bigoni and Drugan (2007) only obtained an \lq optimality condition' for the mismatch.

The homogenization procedure is described in the following three steps, where reference is made to a generic RVE, although results
will be presented for a diluted distribution of randomly located inclusions.

\begin{description}

\item[Step 1.] Consider a RVE made up of a heterogeneous Cauchy material (C), Fig. \ref{step1fig} (left), occupying a region  $$
\Omega_{RVE}^C\equiv\Omega_{1}^C\cup\Omega_{2}^C,
$$
where an inclusion, phase \lq $2$' (occupying the region $\Omega_{2}^C$ and with elastic tensor $\capC^{(2)}$), is fully enclosed
in a matrix, phase \lq $1$' (occupying the region $\Omega_{1}^C$ and with elastic tensor
$\capC^{(1)}$), so that the constitutive local tensor $\capC(\bx)$
within the RVE can be defined as the piecewise constant function
\beq
\capC(\bx)=\begin{cases}\capC^{(1)}~~~\bx \in \Omega_{1}^C, \\ \\ \capC^{(2)}~~~\bx \in \Omega_{2}^C,\end{cases}
\eeq
and the volume fraction $f$ of the inclusion phase can be defined as
\beq\label{volumeratio}
f=\ds\frac{\ds\Omega_{2}^C}{\ds\Omega_{RVE}^C}.
\eeq
The equivalent material is a homogeneous SGE material, Fig. \ref{step1fig} (right),
occupying the region $\Omega_{eq}^{SGE}$
\beq
\Omega_{eq}^{SGE}=\Omega_{RVE}^{C},
\eeq
and constitutive elastic tensors $\capC^{eq}$ (local part) and $\capA^{eq}$ (nonlocal part).
Since the region $\Omega_{eq}^{SGE}$ of the equivalent SGE material  corresponds by definition
to the region $\Omega_{RVE}^C$ of the heterogeneous RVE, in the following both these domains may be
 identified as $\Omega$.

\begin{figure*}[!htcb]
  \begin{center}
\includegraphics[width=9 cm]{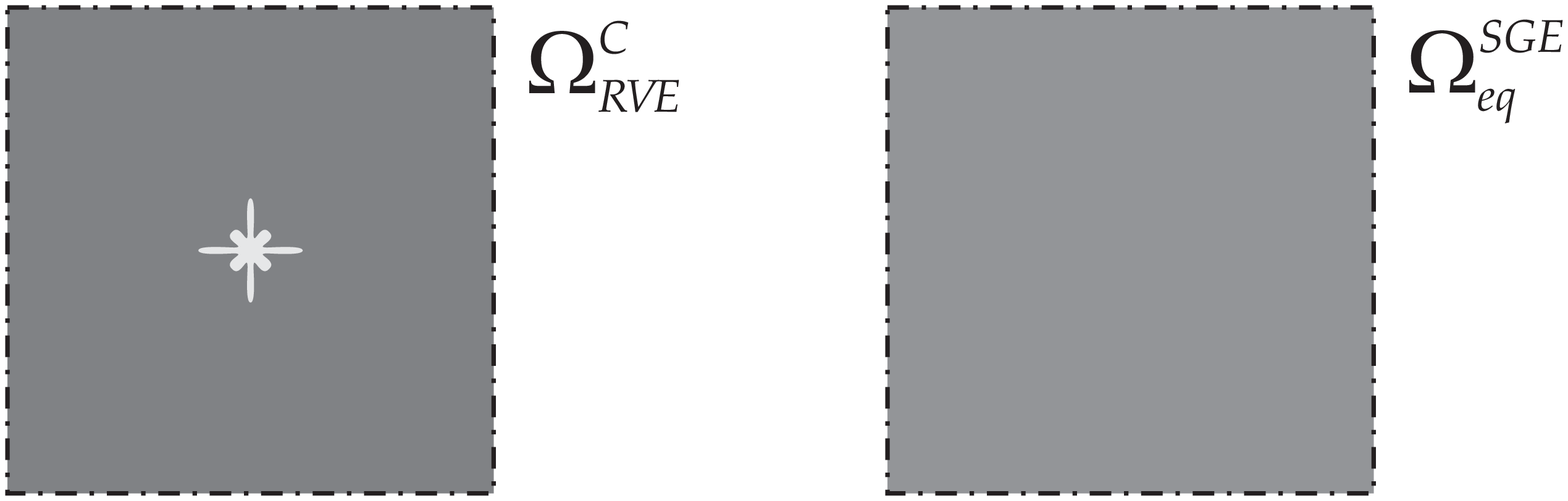}
\caption{\footnotesize Left: Heterogeneous Cauchy-elastic RVE where a matrix of elastic tensor $\capC^{(1)}$ contains a
generic inclusion of elastic tensor $\capC^{(2)}$. Right:
Homogeneous equivalent SGE material with local tensor $\capC^{eq}$ and nonlocal tensor $\capA^{eq}$.}
 \label{step1fig}
 \end{center}
\end{figure*}

\item[Step 2.]

Impose on the RVE  boundary the following second-order (linear and quadratic) displacement field $\overline{\bu}$, Fig. \ref{bcs} (left)
\beq
\lb{boundaryconditions_aa}
\bu=\overline{\bu}, ~~~ \mbox{on } \partial\Omega_{RVE}^C,
\eeq
with
\beq\lb{displacements}
\overline{u}_i=\underbrace{\alpha_{ij} x_j}_{\ds \overline{u}_i^\alpha}+\underbrace{\beta_{ijk} x_j x_k}_{\ds \overline{u}_i^\beta},
\eeq
where $\alpha_{ij}$ and $\beta_{ijk}$ are constant coefficients, the latter having the
symmetry $\beta_{ijk}$=$\beta_{ikj}$.

Impose on the equivalent homogeneous SGE boundary
again the displacement (\ref{displacements}), but together with its normal derivative, Fig. \ref{bcs} (right), so that
\beq
\lb{boundaryconditions2}
\left\{
\begin{array}{lll}
\bu=\overline{\bu}, \\[4mm]
D \bu=D\overline{\bu},
\end{array}
\right.
~~~ \mbox{on } \partial\Omega_{eq}^{SGE}.
\eeq
Note that the mean value of the local strain gradient, which cannot be controlled solely by Dirichlet conditions, is defined
by {\it imposing} the Neumann condition (\ref{boundaryconditions2})$_2$.
This condition can be justified through consideration of the dilute assumption, so that
the influence of the inclusion on the normal derivative is  negligible near the boundary of the RVE.

The imposition of the boundary conditions (\ref{boundaryconditions_aa}) on the RVE
and (\ref{boundaryconditions2}) on the equivalent SGE corresponds, respectively, to the two strain energies
\beq
\lb{strain energies}
\mathcal{W}_{RVE}^C=\ds
\int_{\Omega_{1}^C} \left. w^C\right|_{\capC^{(1)}}+
\int_{\Omega_{2}^C} \left.w^C\right|_{\capC^{(2)}}, \qquad
\mathcal{W}_{eq}^{SGE}=\int_{\Omega_{eq}^{SGE}} \left.w^{SGE}\right|_{\capC^{eq},\capA^{eq}},
\eeq
so that for a generic quadratic displacement field, eqn. (\ref{displacements}),
an energy mismatch (or \lq gap') $\mathcal{G}$ between the two materials
arises as a function of the unknown equivalent constitutive tensor $\capA^{eq}$
\beq
\lb{gapeqn}
\mathcal{G}\left(\capC^{(1)},\capC^{(2)},\capC^{eq},\capA^{eq}\right) =
\mathcal{W}_{RVE}^{C}-
\mathcal{W}_{eq}^{SGE}.
\eeq

\begin{figure*}[!htcb]
  \begin{center}
\includegraphics[width=10 cm]{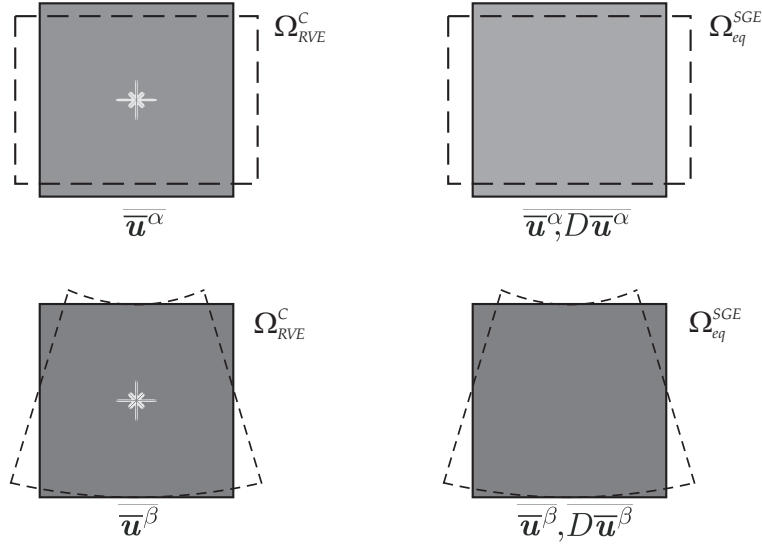}
\caption{\footnotesize Imposition of the same linear (top) and quadratic (bottom) boundary displacement conditions on the  heterogeneous Cauchy RVE (left) and
on the homogeneous equivalent SGE (right).
In the homogeneous equivalent SGE (right) the normal derivative of displacement (Neumann condition) is also imposed at the boundary.}
 \label{bcs}
 \end{center}
\end{figure*}

\item[Step 3.] Find the unknown equivalent constitutive tensor $\capA^{eq}$ by imposing a null
energy mismatch $\mathcal{G}$
\beq
\lb{minimization}
\mathcal{G}\left(\capC^{(1)},\capC^{(2)},\capC^{eq},\capA^{eq}\right) =
0.
\eeq

Note that in the case of purely linear displacements ($\bbeta=\b0$) the energy mismatch $\mathcal{G}$
 is null by definition of
$\capC^{eq}$. On the other hand, when quadratic displacements are considered, an energy mismatch $\mathcal{G}$
is different from  zero and it can be tuned
to vanish by changing the value of the unknown tensor $\capA^{eq}$.
\end{description}

The above-procedure is general, but subsequent calculations will be limited to the dilute approximation, and the results will be an extension
of Bigoni and Drugan (2007) since (i.) the inclusions are of arbitrary shape and, more interestingly, (ii.) the comparison material,
a Mindlin elastic second-gradient material, allows a perfect match of  the energies (while Bigoni and Drugan (2007) did consider only cylindrical
or spherical inclusions and  were only able to provide a minimization of energy gap).

\subsection{Assumptions about geometrical properties of matrix and inclusion phases}

Henceforth the following geometrical properties for both the subsets $\Omega_{1}^C$and $\Omega_{2}^C$ will be assumed:\footnote{
Note that, by definition of static moment vector $\bS$ and Euler tensor of inertia $\bE$, eqn (\ref{euler}),
the geometrical properties \textbf{GP1}, eqn (\ref{Eulerphases}) and \textbf{GP2}, eqn (\ref{Eulerphases2}),
of the subsets $\Omega_{1}^C$and $\Omega_{2}^C$
are also necessarily satisfied by $\Omega_{RVE}^C$, so that
\beq
\lb{Eulerrve}
\bS(\Omega_{RVE}^C)=\b0, \qquad
\bE(\Omega_{RVE}^C)= \rho^2 \Omega_{RVE}^C\Id,
\eeq
where the radius $\rho=\rho(\Omega_{RVE}^C)$  is related to the radii of the matrix $\rho^{(1)}$ and the inclusion $\rho^{(2)}$
as follows
\beq\label{legamerho}
\rho^2=(1-f)\left[\rho^{(1)}\right]^2+f\left[\rho^{(2)}\right]^2.
\eeq
}
\begin{description}
\item[GP1)]
The centroids of the matrix and of the inclusion coincide and correspond to the origin of the $x_i$--axes,
so that both the static moments of the inclusion and of the matrix are null
\beq
\lb{Eulerphases}
\bS(\Omega_{1}^C)=\b0, \qquad
\bS(\Omega_{2}^C)=\b0.
\eeq

\item[GP2)] The $x_i$--axes are principal axes of inertia for both the matrix and the inclusion
and the ellipsoids of inertia are a sphere (or a circle in 2D)
\beq
\lb{Eulerphases2}
\bE(\Omega_{1}^C) = \left[\rho^{(1)}\right]^2 \Omega_{1}^C\Id, \qquad
\bE(\Omega_{2}^C)= \left[\rho^{(2)}\right]^2 \Omega_{2}^C\Id,
\eeq
where $\Id$ is the identity second-order tensor and the second-order Euler tensor of inertia $\bE$ relative to the $x_i$--axes,
defined for a generic solid occupying the region $V$ as
\beq\label{euler}
E_{ij}(V) =\int_{V} x_i\,x_j,
\eeq
while $\rho^{(1)}=\rho(\Omega_{1}^C)$ and $\rho^{(2)}=\rho(\Omega_{2}^C)$ are the radii
of the spheres (or circles in 2D) of inertia of the matrix and the inclusion.
Note that the assumption of spherical tensors of inertia yields a spherical tensor for the RVE,
which is coherent with the assumption of randomness of the distribution of inclusions.

\item[GP3)] The radius of the sphere of inertia for the inclusion phase vanishes in the limit of null inclusion volume fraction
\beq\label{gp3}
\lim_{f\rightarrow0}\rho^{(2)}(f)=0,
\eeq
or, equivalently, all the dimensions of the inclusion (and therefore
 the radius of the smallest ball containing the inclusion) are zero for $f=0$.
\end{description}

\begin{figure*}[!htcb]
  \begin{center}
\includegraphics[width=10 cm]{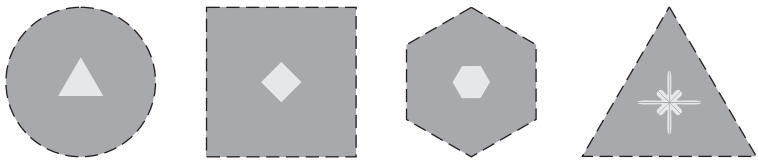}
\caption{\footnotesize Some examples of two-dimensional RVE  satisfying the geometrical properties
\textbf{GP1}, eqn (\ref{Eulerphases}), and \textbf{GP2}, eqn (\ref{Eulerphases2}), for plane strain condition.}
 \label{gp2_example}
 \end{center}
\end{figure*}

\begin{figure*}[!htcb]
  \begin{center}
\includegraphics[width=14 cm]{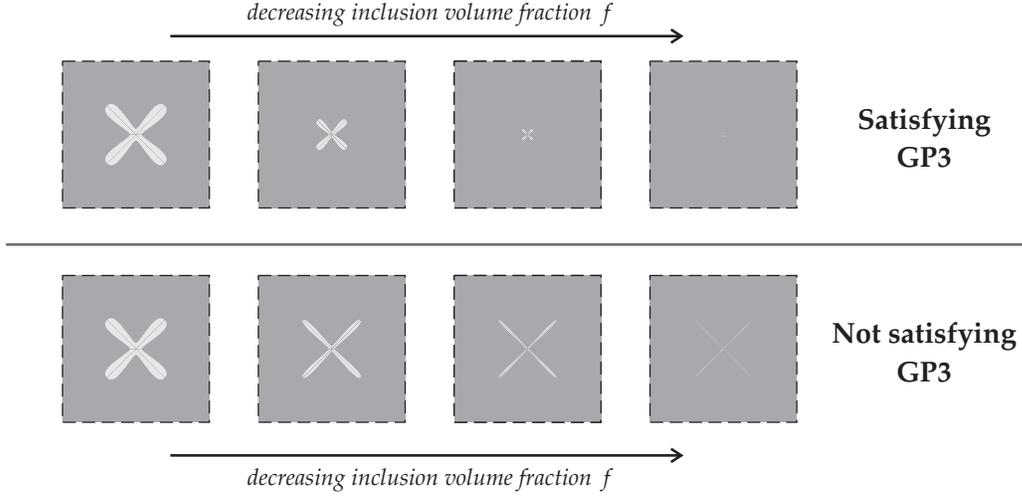}
\caption{\footnotesize Examples of two-dimensional RVE satisfying (upper part)
or not (lower part) the geometrical property \textbf{GP3}, eqn (\ref{gp3}).
In the lower part, the radius of inertia of the inclusion does not vanish in the limit of vanishing volume fraction.}
 \label{gp3_example}
 \end{center}
\end{figure*}

Examples of two-dimensional  RVE, characterized by the geometrical properties
\textbf{GP1}-\textbf{GP2} and \textbf{GP3}
are reported in Figs. \ref{gp2_example} and \ref{gp3_example}, respectively.

\section{Equivalent nonlocal properties from homogenization in the dilute case}

The following proposition is the central result in this article, providing {\it the nonlocal effective tensor from
second-order homogenization of a heterogeneous Cauchy RVE
containing a small inclusion}.

\paragraph{Homogenization proposition.} For a dilute concentration of the inclusion phase ($f\ll1$) and assuming the geometrical properties
\textbf{GP1} - \textbf{GP2} - \textbf{GP3} for the RVE, the nonlocal sixth-order tensor $\capA^{eq}$ of the equivalent SGE material is evaluated (at first-order in $f$) as
\beq
\lb{sol}
\begin{split}
\capA^{eq}_{ijhlmn}&=-f \frac{\rho^2}{4}\left(
\tilde{\capC}_{ihln}\delta_{jm}+
\tilde{\capC}_{ihmn}\delta_{jl}+
\tilde{\capC}_{jhln}\delta_{im}+
\tilde{\capC}_{jhmn}\delta_{il}
\right) + o(f),
\end{split}
\eeq
where $\rho$ is the radius of the sphere (or circle in 2D) of inertia of the RVE cell, and
$\tilde{\capC}$ is introduced to define (at first-order in $f$) the difference  between
the local constitutive tensors for the effective material $\capC^{eq}$ and the matrix $\capC^{(1)}$, so that
\beq
\label{valtari}
\capC^{eq}=\capC^{(1)}+f\tilde{\capC},
\eeq
which is assumed to be known from standard homogenization, performed on linear displacement boundary conditions.

Eqn (\ref{sol}) represents the solution of the homogenization problem and is obtained by imposing
the vanishing of the energy mismatch $\mathcal{G}$, eqn (\ref{minimization}),
when the same second-order  displacement boundary conditions are applied both on the
heterogeneous Cauchy material and on the homogeneous equivalent SGE material,
eqns (\ref{boundaryconditions_aa}) and (\ref{boundaryconditions2}), respectively.

From the solution (\ref{sol}),
in agreement with Bigoni and Drugan (2007), it can be noted that:
\begin{itemize}
\item the equivalent SGE material is positive definite if and only if $\tilde{\capC}$ is negative definite;
\item the constitutive higher-order tensor $\capA^{eq}$ is linear in $f$ for dilute concentration.
\end{itemize}

\paragraph{Proof of the homogenization proposition}
\begin{enumerate}[i)]
\item Consider the second-order (linear and quadratic) displacement boundary condition (\ref{boundaryconditions2}) applied
on the boundary of a homogeneous SGE material with constitutive tensors $\capC$ and $\capA$.
In the absence of body force, $\bb=\b0$, let us consider
 the extension within the body of the quadratic displacement field $\overline{\bu}$, eqn (\ref{displacements}), applied on the boundary
\beq
\lb{quadraticdisplacementinbody}
u_i=\underbrace{\alpha_{ij} x_j}_{\ds u_i^\alpha}+\underbrace{\beta_{ijk} x_j x_k}_{\ds u_i^\beta},
\qquad \bx \, \mbox{in} \, \Omega,
\eeq
providing the following deformation $\bvarepsilon$ and curvature $\bchi$ fields
\beq
\lb{straincurvaturefield}
\varepsilon_{ij}=\frac{\alpha_{ij}+\alpha_{ji}}{2}
+(\beta_{ijk}+\beta_{jik})x_k, \qquad
\chi_{ijk}=2\beta_{kij},
\eeq
and the following stress $\bsigma$ and double-stress $\btau$ fields,
\beq\lb{stresscurvaturefield}
\begin{split}
\sigma_{ij}=\capC_{ijhk}\alpha_{hk}
+2\capC_{ijhk}\beta_{hkl}x_l,\qquad
\tau_{ijk}=2\capA_{ijklmn}\beta_{nlm}.
\end{split}
\eeq
The  stress field (\ref{stresscurvaturefield}) follows from the displacement field (\ref{quadraticdisplacementinbody}) and satisfies
the equilibrium equation (\ref{indequilibrium}) if and only if \footnote{Note that the constraint (\ref{diamondrestraint})
arises independently of whether the material is Cauchy elastic or SGE.}
\beq\label{diamondrestraint}
\capC_{ijhk}\beta_{hkj}=\b0,
\eeq
which for  isotropic homogeneous materials reduces to the condition obtained by Bigoni and Drugan (2007)
\beq
\lb{condbeta}
\beta_{jji}=-(1-2\nu)\beta_{ikk},
\eeq
(with Poisson's ratio $\nu=\lambda/2(\lambda+\mu)$).

In the following
we will use the superscript $^{\diamond}$ for $\bbeta$ (namely, $\bbeta^{\diamond}$) to denote the components of the  third-order tensor
$\bbeta$
satisfying eqn (\ref{diamondrestraint}), or (\ref{condbeta}) for isotropy.

\item Consider an auxiliary material with local constitutive tensor
$\capC^*$, defined as a first-order perturbation in $f$ to the equivalent local constitutive tensor $\capC^{eq}$, namely,
\beq
\label{CauxiliarySGE}
\capC^*=\capC^{eq}+f\left(\hat{\capC}-\tilde{\capC}\right),
\eeq
so that using eqn (\ref{valtari}) we can write
\beq
\lb{Cauxiliary0}
\capC^* = \capC^{(1)}+ f \hat{\capC},
\eeq
where $\hat{\capC}$, together with $\capC^*$, define an arbitrary material with properties \lq close' to both the matrix and the equivalent material, an arbitrariness which will be used later to
eliminate the constraint (\ref{diamondrestraint}).
By definition, the displacement field
\beq
\lb{quadraticdisplacementinbodyasterisk}
u^*_i=\underbrace{\alpha_{ij} x_j}_{\ds u_i^\alpha}+\underbrace{\beta_{ijk}^{\diamond *} x_j x_k}_{\ds u_i^{\beta^{\diamond *}}},
\qquad \bx \, \mbox{in} \, \Omega.
\eeq
is equilibrated [in other words satisfies eqn (\ref{diamondrestraint})] in a homogeneous material characterized by the constitutive tensor $\capC^*$ and it corresponds to the following quadratic displacement field on the boundary
\beq
\lb{quadraticdisplacementasterisk}
\overline{u}^*_i=\underbrace{\alpha_{ij} x_j}_{\ds \overline{u}_i^\alpha}+
\underbrace{\beta_{ijk}^{\diamond *} x_j x_k}_{\ds\overline{u}_i^{\beta^{\diamond *}}},
\qquad \bx \, \mbox{on} \, \partial\Omega.
\eeq

\item Apply  on the boundary $\partial  \Omega_{RVE}^{C}$ of the heterogeneous Cauchy material (RVE)
the displacement boundary condition (\ref{quadraticdisplacementasterisk}),
\beq\label{bcrve}
\overline{\bu}^{RVE}=\overline{\bu}^*,
\qquad \mbox{on} \, \partial\Omega_{RVE}^C.
\eeq

According to {\bf Lemma 1} (Appendix \ref{lemma1subsection}), the strain energy in the RVE
at first-order in $f$ is the sum of the strain energy due to the linear ($\balpha$) and nonlinear ($\bbeta$) displacement boundary conditions,
and the mutual strain energy, say, the \lq $\balpha-\bbeta$ energy term' is null at first-order in $f$,\footnote{
Considering that the RVE satisfies geometrical symmetry conditions,
in addition to the geometrical properties \textbf{GP1} and \textbf{GP2},
it can be proven that the mutual energy is identically null even in the case of non-dilute suspension of inclusion
 \beq
\mathcal{W}_{RVE}^{C}\left(\overline{\bu}^*\right)=
\mathcal{W}_{RVE}^{C}\left(\overline{\bu}^{\alpha}\right)+
\mathcal{W}_{RVE}^{C}\left(\overline{\bu}^{\beta^{\diamond *}}\right), \qquad \forall \, f.
\eeq}
so that
\beq\label{energy1rve}
\mathcal{W}_{RVE}^{C}\left(\overline{\bu}^*\right)=
\mathcal{W}_{RVE}^{C}\left(\overline{\bu}^{\alpha}\right)+
\mathcal{W}_{RVE}^{C}\left(\overline{\bu}^{\beta^{\diamond *}}\right) + o(f).
\eeq

\item Apply on the boundary $\partial  \Omega_{eq}^{SGE}$ of the homogeneous SGE material
 the same displacement boundary condition $\overline{\bu}^*$, eqn (\ref{quadraticdisplacementasterisk}),
 imposed to the RVE and  complemented
by the higher-order boundary condition in terms of displacement normal derivative
taken equal\footnote{The displacement
field eqn (\ref{quadraticdisplacementinbodyasterisk}) is the solution for
a homogeneous SGE  when boundary conditions  (\ref{bcsge}) are imposed.
It can be easily proven that the result of the proposed homogenization procedure holds when the
higher-order boundary condition changes as $\overline{D\bu}^{SGE}=D\overline{\bu}^{RVE}$ since
the strain energy developed in the SGE material is the same at the first order
$$
\mathcal{W}_{eq}^{SGE}\left(\overline{\bu}^*, D\overline{ \bu}^{RVE}\right)=
\mathcal{W}_{eq}^{SGE}\left(\overline{\bu}^*, D\overline{\bu}^{*}\right)+o(f).
$$} to  $D\overline{\bu}^*$
\beq\label{bcsge}
\left\{
\begin{array}{ll}
\overline{\bu}^{SGE}= \overline{\bu}^{*},\\[3 mm]
\overline{D \bu}^{SGE}= D \overline{\bu}^{*},
\end{array}
\right.
 \qquad \mbox{on} \,\partial  \Omega_{eq}^{SGE},
\eeq
where $D \overline{\bu}^{*}$ is the normal derivative of the displacement field (\ref{quadraticdisplacementinbodyasterisk}).

According to the result presented in {\bf Lemma 2} (Appendix \ref{lemma2subsection}),
the $\balpha-\bbeta$ energy term
is null and
the strain energy in $\Omega_{eq}^{SGE}$ is
\beq
\label{energy1sge}
\mathcal{W}_{eq}^{SGE}\left(\overline{\bu}^*, D\overline{\bu}^{*}\right)=
\mathcal{W}_{eq}^{SGE}\left(\overline{\bu}^{\alpha}, D\overline{\bu}^{\alpha}\right)+
\mathcal{W}_{eq}^{SGE}\left(\overline{\bu}^{\beta^{\diamond *}}, D\overline{ \bu}^{\beta^{\diamond *}}\right),
\eeq
where $D\overline{\bu}^{\alpha}$ and $D\overline{\bu}^{\beta^{\diamond *}}$ are the contributions of the imposed normal derivative
depending on $\balpha$ and $\bbeta$ terms in $D \overline{\bu}^{*}$, respectively.

\item The energy minimization procedure, eqn (\ref{minimization}), can be performed using
the energy stored in the heterogeneous Cauchy material $\mathcal{W}_{RVE}^{C}$, eqn (\ref{energy1rve}),
and in the homogeneous SGE material $\mathcal{W}_{eq}^{SGE}$, eqn (\ref{energy1sge}), so that the energy mismatch is given by
\beq
\mathcal{G}\left(\capC^{(1)},\capC^{(2)},\capC^{eq},\capA^{eq}\right) =
\mathcal{G}^\alpha\left(\capC^{(1)},\capC^{(2)},\capC^{eq},\capA^{eq}\right)+
\mathcal{G}^{\beta^{\diamond *}}\left(\capC^{(1)},\capC^{(2)},\capC^{eq},\capA^{eq}\right)
\eeq
where
\beq\begin{array}{ll}\label{rice}
\mathcal{G}^\alpha\left(\capC^{(1)},\capC^{(2)},\capC^{eq},\capA^{eq}\right)=&\mathcal{W}_{RVE}^{C}\left(\overline{\bu}^{\alpha}\right)
-\mathcal{W}_{eq}^{SGE}\left(\overline{\bu}^{\alpha}, D\overline{ \bu}^{\alpha}\right),\\[5mm]
\mathcal{G}^{\beta^{\diamond *}}\left(\capC^{(1)},\capC^{(2)},\capC^{eq},\capA^{eq}\right)=&\mathcal{W}_{RVE}^{C}\left(\overline{\bu}^{\beta^{\diamond *}}\right)
-\mathcal{W}_{eq}^{SGE}\left(\overline{\bu}^{\beta^{\diamond *}}, D\overline{\bu}^{\beta^{\diamond *}}\right).
\end{array}
\eeq
Since only the local contribution (depending on $\capC^{eq}$) arises in the SGE strain energy when the
linear boundary displacement condition
($\bbeta^{\diamond *}=\b0$ and $\overline{\bu}^{SGE}=\overline{\bu}^{\alpha}$, $\overline{D \bu}^{SGE}=D\overline{ \bu}^{\alpha}$)
 is imposed
(while the non-local contribution depending on $\capA^{eq}$ is identically null because higher-order stress and curvature are null),
the energy mismatch $\mathcal{G}^\alpha$ due to the $\alpha$ terms is null by definition of $\capC^{eq}$
(which is known from the first-order homogenization procedure)
\beq\begin{array}{ll}\label{galfaalfa}
\mathcal{G}^\alpha\left(\capC^{(1)},\capC^{(2)},\capC^{eq},\capA^{eq}\right)=&\mathcal{G}^\alpha\left(\capC^{(1)},\capC^{(2)},\capC^{eq}\right)=0.
\end{array}
\eeq
Therefore, the proposed energy minimization procedure, based on linear and quadratic displacement boundary
condition and leading to the definition of $\capA^{eq}$,
can be performed referring only to the $\beta^{\diamond *}$ terms,
\beq
\label{cranberry}
\mathcal{G}\left(\capC^{(1)},\capC^{(2)},\capC^{eq},\capA^{eq}\right) =
\mathcal{G}^{\beta^{\diamond *}}\left(\capC^{(1)},\capC^{(2)},\capC^{eq},\capA^{eq}\right).
\eeq

\item Keeping into account the results presented in {\bf Lemma 3} (Appendix \ref{lemma3subsection}) and {\bf Lemma 4} (Appendix \ref{lemma4subsection}), the energy mismatch (\ref{cranberry}) is given by the difference of the following two terms
\beq
\lb{jonsi}
\mathcal{W}_{RVE}^{C}(\overline{\bu}^{\beta^{\diamond *}})=
2\rho^2 \Omega\capC^{(1)}_{ijhk}\beta^{\diamond *}_{ijl}\beta^{\diamond *}_{hkl}+o(f).
\eeq
and
\beq
\lb{alex}
\mathcal{W}_{eq}^{SGE}(\overline{\bu}^{\beta^{\diamond *}},D\overline{\bu}^{\beta^{\diamond *}})=
2 \Omega\left(\rho^2\capC^{eq}_{ijhk}\delta_{lm}+\capA^{eq}_{jlikmh}\right)\beta^{\diamond
*}_{ijl}\beta^{\diamond *}_{hkm}+o(f).
\eeq

\item Therefore, from eqns (\ref{valtari}), (\ref{jonsi}) and (\ref{alex}),
the annihilation of the strain energy gap $\mathcal{G}$, eqn (\ref{cranberry}) (between the real heterogeneous Cauchy
and the equivalent homogeneous SGE materials) is represented by the condition
\beq
\lb{annihilationG}
\left(f\rho^2\tilde{\capC}_{ijhk}\delta_{lm}+
\capA^{eq}_{jlikmh}\right)\beta^{\diamond *}_{ijl}\beta^{\diamond *}_{hkm}+o(f)=0.
\eeq

\item The energy annihilation (\ref{annihilationG}) has been obtained for a nonlinear displacement field $\bbeta^{\diamond *}$, in equilibrium within
a homogeneous material with local constitutive tensor $\capC^*$.
But, according to eqn (\ref{Cauxiliary0}), tensor $\capC^*$ defines an arbitrary material, so that using this arbitrariness we obtain
\beq
\lb{annihilationGgeneric}
\left(f\rho^2\tilde{\capC}_{ijhk}\delta_{lm}+
\capA^{eq}_{jlikmh}\right)\beta_{ijl}\beta_{hkm}
+o(f) =0,
\eeq
where the components of $\bbeta$ are unrestricted, except  for the symmetry $\beta_{ijk}$=$\beta_{ikj}$.
Eventually, the annihilation of energy mismatch $\mathcal{G}$, eqn (\ref{annihilationGgeneric}),
defines the non-local constitutive tensor $\capA^{eq}$ for the equivalent SGE material as in
eqn (\ref{sol}).
$\Box$

\end{enumerate}

\section{Conclusions}
Micro- or nano-structures embedded in solids introduce internal length-scales and nonlocal effects within
the mechanical modelling, leading to  higher-order
theories. We have provided an analytical approach to the determination of
the parameters defining an elastic higher-order (Mindlin) material, as
the homogenization of a heterogeneous Cauchy elastic material, eqn (\ref{sol}). This
result, obtained through the proposed homogenization procedure, is limited
to the dilute approximation, but is not
restricted to isotropy of the constituents and leaves a certain freedom to
the shape of the inclusions. A perfect match between the elastic
energies of the heterogeneous and homogeneous materials is obtained.
Examples and results on material symmetry and positive definiteness are deferred to part II of this article (Bacca et al., 2013).

\vspace*{5mm} \noindent {\sl Acknowledgments}
M. Bacca gratefully acknowledges financial support from Italian Prin 2009 (prot. 2009XWLFKW-002).
D. Bigoni, F. Dal Corso and D. Veber  gratefully acknowledge financial support from the grant PIAP-GA-2011-286110-INTERCER2,
\lq Modelling and optimal design of ceramic structures with defects and imperfect interfaces'.
\vspace*{10mm}

 { \singlespace
}

\appendix

\section{Proofs of lemmas 1-4}

\setcounter{equation}{0}
\renewcommand{\theequation}{{A}.\arabic{equation}}
\subsection{Lemma 1: Null mutual $\balpha$--$\bbeta$ energy term  for the RVE
at the first-order in concentration $f$}\label{lemma1subsection}

\paragraph{Statement.}
When a quadratic displacement $\overline{\bu}^{*}$, eqn (\ref{quadraticdisplacementasterisk}), is applied on the boundary of
a RVE satisfying the geometrical property \textbf{GP1} and \textbf{GP3},
the strain energy at first-order in $f$ is given by eqn (\ref{energy1rve}).
\paragraph{Proof.}
By the superposition principle, the fields originated by the application of
$\overline{\bu}^*=\overline{\bu}^\alpha+\overline{\bu}^{\beta^{\diamond*}}$ are given by
the sum of the respective fields originated from the boundary conditions $\overline{\bu}^\alpha$
and $\overline{\bu}^{\beta^{\diamond*}}$
\beq\lb{decompositionrve}
\bvarepsilon(\bx)=\bvarepsilon^\alpha(\bx)+\bvarepsilon^{\beta^{\diamond*}}(\bx),\qquad\;\;
\bsigma(\bx)=\bsigma^\alpha(\bx)+\bsigma^{\beta^{\diamond*}}(\bx),
\eeq
(the latter calculated through the constitutive eqn (\ref{defconstsgm})$_1$) so that the strain energy (\ref{strain energies})$_1$ becomes
\beq
\mathcal{W}^C_{RVE}(\overline{\bu}^*)=
\mathcal{W}^C_{RVE}(\overline{\bu}^\alpha)+
\mathcal{W}^C_{RVE}(\overline{\bu}^{\beta^{\diamond*}})+
\underbrace{\mathcal{W}^C_{RVE}(\overline{\bu}^\alpha;\overline{\bu}^{\beta^{\diamond*}})}_{mutual\,\,energy}
\eeq
where
\beq\label{energiesRVE}
\begin{array}{lll}
\ds\mathcal{W}^C_{RVE}(\overline{\bu}^\alpha)=\frac{1}{2}
\int_{\Omega_R}\varepsilon^{\alpha}_{ij}(\bx) \capC_{ijhk}(\bx)\varepsilon^{\alpha}_{hk}(\bx),\\[6mm]
\ds\mathcal{W}^C_{RVE}(\overline{\bu}^{\beta^{\diamond*}})=\frac{1}{2}
\int_{\Omega_R}\varepsilon^{\beta^{\diamond*}}_{ij}(\bx) \capC_{ijhk}(\bx)\varepsilon^{\beta^{\diamond*}}_{hk}(\bx),\\[6mm]
\ds\mathcal{W}^C_{RVE}(\overline{\bu}^\alpha;\overline{\bu}^{\beta^{\diamond*}})=
\int_{\Omega_R}\varepsilon^{\alpha}_{ij}(\bx) \capC_{ijhk}(\bx)\varepsilon^{\beta^{\diamond*}}_{hk}(\bx).
\end{array}
\eeq

Through two applications of the principle of virtual work\footnote{
In the first application, the fields corresponding to the solution (\ref{decompositionrve}) are considered
\beq
\int_{\Omega_R}\varepsilon^{\alpha}_{ij}(\bx) \sigma^{\beta^{\diamond*}}_{ij}(\bx)=
\int_{\partial\Omega_R}\overline{u}_i^\alpha(\bx) t^{\beta^{\diamond*}}_{i}(\bx),
\eeq
while in the second application, the kinematical field generated by the admissible
displacement $\bu^\alpha$ (\ref{quadraticdisplacementinbodyasterisk})
within the RVE is considered so that the mutual energy
(\ref{mutualmutual}) is obtained.
}
the mutual energy (\ref{energiesRVE})$_3$ can be computed as
\beq\label{mutualmutual}
\ds\mathcal{W}^C_{RVE}(\overline{\bu}^\alpha;\overline{\bu}^{\beta^{\diamond*}})=
\alpha_{ij}\int_{\Omega_R}\sigma^{\beta^{\diamond*}}_{ij}(\bx),
\eeq
which, using the constitutive relation (\ref{defconstsgm})$_1$ and the symmetries of the local constitutive tensors $\capC^{(1)}$ and $\capC^{(2)}$,
can be decomposed as the sum of two contributions
\beq\lb{mutual4}
\ds\mathcal{W}^C_{RVE}(\overline{\bu}^\alpha;\overline{\bu}^{\beta^{\diamond*}})=
\alpha_{ij}\capC^{(1)}_{ijhk}\,\int_{\Omega_{R}}u^{\beta^{\diamond*}}_{h,k}(\bx)
+\alpha_{ij}\left(\capC^{(2)}_{ijhk}-\capC^{(1)}_{ijhk}\right)\,\int_{\Omega_{R_2}}u^{\beta^{\diamond*}}_{h,k}(\bx).
\eeq

Through two further applications of the divergence theorem and using the geometrical property \textbf{GP1} for the RVE,\footnote{
In the first application of the divergence theorem,
$u^{\beta^{\diamond*}}=\overline{u}^{\beta^{\diamond*}}$, eqn (\ref{quadraticdisplacementasterisk}),
is considered on the boundary $\partial\Omega_{R}$, so that
\beq\label{green1}
\ds\int_{\Omega_{R}}u^{\beta^{\diamond*}}_{h,k}(\bx)=\beta^{\diamond*}_{hlm}\int_{\partial\Omega_{R}}n_kx_l x_m,
\eeq
while, in the second application,
the kinematically admissible displacement field
 $\overline{u}^{\beta^{\diamond*}}$, eqn (\ref{quadraticdisplacementinbodyasterisk}), is considered within the RVE, yielding
\beq\label{green2}
\ds \beta^{\diamond*}_{hlm}\int_{\partial\Omega_{R}}n_kx_l x_m=2\beta^{\diamond*}_{hlk}\int_{\Omega_{R}}x_l,
\eeq
so that the geometrical property \textbf{GP1} for the RVE leads to eqn (\ref{firsttermlemma1}).}
the first term on the right-hand-side of eqn (\ref{mutual4}) results to be null
\beq\label{firsttermlemma1}
\alpha_{ij}\capC^{(1)}_{ijhk}\int_{\Omega_{R}}u^{\beta^{\diamond*}}_{h,k}(\bx)=0.
\eeq
Introducing the mean value over a domain $\Omega$ of the function $f(\bx)$ as
\beq
\left.\left<f(\bx)\right>\right|_{\Omega}=\frac{1}{\Omega}\int_{\Omega}f(\bx),
\eeq
the second term on the right-hand-side of eqn (\ref{mutual4}) can be rewritten as
\beq\label{nonloso}
\alpha_{ij}\left(\capC^{(2)}_{ijhk}-\capC^{(1)}_{ijhk}\right)\,\Omega_{R_2}
\left.\left<u^{\beta^{\diamond*}}_{h,k}(\bx)\right>\right|_{\Omega_{R_2}}.
\eeq
Assuming the geometrical property \textbf{GP3} for the RVE,
the displacement field in the presence of the inclusion is given by the asymptotic expansion in the volume fraction $f$
\beq\lb{mutual5}
u_i^{\beta^{\diamond*}}=\beta^{\diamond*}_{ijk}x_jx_k+f^q\,\tilde{u}_{i}^{\beta^{\diamond*}}+o(f),
\eeq
subject to the constraint
\beq\label{qu}
0<q\leq1,
\eeq
and considering the geometrical property \textbf{GP1} for the RVE,
 together with the definition of volume fraction $f$, eqn (\ref{volumeratio}),
expression (\ref{nonloso}) becomes
\beq\lb{mutual6}
f^{q+1}\Omega\,\alpha_{ij}\left(\capC^{(2)}_{ijhk}-\capC^{(1)}_{ijhk}\right)
\left.\left<\tilde{u}^{\beta^{\diamond*}}_{h,k}(\bx)\right>\right|_{\Omega_{R_2}},
\eeq
from which, considering the restriction on the power $q$ (\ref{qu}), the second term on the
right-hand-side of eqn (\ref{mutual4}) is null at first-order in $f$
\beq\label{secondtermlemma1}
\alpha_{ij}\left(\capC^{(2)}_{ijhk}-\capC^{(1)}_{ijhk}\right)\,\int_{\Omega_{R_2}}u^{\beta^{\diamond*}}_{h,k}(\bx)=o(f).
\eeq
Considering results (\ref{firsttermlemma1}) and (\ref{secondtermlemma1}), the mutual energy in the RVE (\ref{energiesRVE})$_3$
is null at first-order in $f$ and proposition (\ref{energy1rve}) follows.
$\Box$

\subsection{Lemma 2: Null mutual $\balpha$--$\bbeta$ energy term  for the homogeneous SGE}\label{lemma2subsection}

\paragraph{Statement.}

When a quadratic displacement $\overline{\bu}^{*}$, eqn (\ref{quadraticdisplacementasterisk}),
and  the normal component of its derivative $D\overline{\bu}^*$ are applied on the boundary of
a SGE satisfying the geometrical property \textbf{GP1},
the strain energy is given by eqn (\ref{energy1sge}).
\paragraph{Proof.}
By the superposition principle, the fields originated by the application of
the boundary conditions ($\overline{\bu}^*=\overline{\bu}^\alpha+\overline{\bu}^{\beta^{\diamond*}}$,
$D\overline{\bu}^*=D\overline{\bu}^\alpha+D\overline{\bu}^{\beta^{\diamond*}}$) can be obtained
as the sum of the respective fields arising from the boundary conditions
($\overline{\bu}^\alpha$, $D\overline{\bu}^\alpha$)
and ($\overline{\bu}^{\beta^{\diamond*}}$, $D\overline{\bu}^{\beta^{\diamond*}}$) in the forms
\beq\lb{decompositionsge}
\begin{split}
&\bvarepsilon(\bx)=\bvarepsilon^\alpha(\bx)+\bvarepsilon^{\beta^{\diamond*}}(\bx),\qquad\;\;
\bchi(\bx)=\bchi^\alpha(\bx)+\bchi^{\beta^{\diamond*}}(\bx),\\
&\bsigma(\bx)=\bsigma^\alpha(\bx)+\bsigma^{\beta^{\diamond*}}(\bx),\qquad
\btau(\bx)=\btau^\alpha(\bx)+\btau^{\beta^{\diamond*}}(\bx),
\end{split}
\eeq
(the latter calculated through the constitutive eqn (\ref{defconstsgm}))
so that the strain energy (\ref{strain energies})$_2$ becomes
\beq
\mathcal{W}^{SGE}_{eq}(\overline{\bu}^*,D\overline{\bu}^*)=
\underbrace{\mathcal{W}^{SGE}_{eq}(\overline{\bu}^\alpha,D\overline{\bu}^\alpha)+
\mathcal{W}^{SGE}_{eq}(\overline{\bu}^{\beta^{\diamond*}},D\overline{\bu}^{\beta^{\diamond*}})}_{direct\,\,energy}+
\underbrace{\mathcal{W}^{SGE}_{eq}
(\overline{\bu}^\alpha,D\overline{\bu}^\alpha;\overline{\bu}^{\beta^{\diamond*}},D\overline{\bu}^{\beta^{\diamond*}})}_{mutual\,\,energy}
\eeq
where
\beq\label{energiesSGE}
\begin{array}{lll}
\ds\mathcal{W}^{SGE}_{eq}(\overline{\bu}^\alpha,D\overline{\bu}^\alpha)=\frac{1}{2}
\int_{\Omega_{eq}}\left[\varepsilon^{\alpha}_{ij}(\bx) \capC_{ijhk}^{eq}\varepsilon^{\alpha}_{hk}(\bx)
+\chi^{\alpha}_{ijl}(\bx) \capA_{ijlhkm}^{eq}\chi^{\alpha}_{hkm}(\bx)\right],\\[6mm]
\ds\mathcal{W}^{SGE}_{eq}(\overline{\bu}^{\beta^{\diamond*}},D\overline{\bu}^{\beta^{\diamond*}})=\frac{1}{2}
\int_{\Omega_{eq}}\left[\varepsilon^{\beta^{\diamond*}}_{ij}(\bx) \capC_{ijhk}^{eq}\varepsilon^{\beta^{\diamond*}}_{hk}(\bx)+
\chi^{\beta^{\diamond*}}_{ijl}(\bx) \capA_{ijlhkm}^{eq}\chi^{\beta^{\diamond*}}_{hkm}(\bx)\right],\\[6mm]
\ds\mathcal{W}^{SGE}_{eq}(\overline{\bu}^\alpha,D\overline{\bu}^\alpha;\overline{\bu}^{\beta^{\diamond*}},D\overline{\bu}^{\beta^{\diamond*}})=
\int_{\Omega_{eq}}\left[\varepsilon^{\alpha}_{ij}(\bx) \capC_{ijhk}^{eq}\varepsilon^{\beta^{\diamond*}}_{hk}(\bx)+
\chi^{\alpha}_{ijl}(\bx) \capA_{ijlhkm}^{eq}\chi^{\beta^{\diamond*}}_{hkm}(\bx)\right].
\end{array}
\eeq

Application of the boundary condition $(\overline{\bu}^\alpha, D\overline{\bu}^\alpha)$
 on $\partial\Omega_{eq}$ leads to the displacement field $\bu^\alpha(\bx)$, eqn (\ref{quadraticdisplacementinbodyasterisk}),
 so that $\bchi^\alpha(\bx)=\b0$ and, considering the symmetries of the equivalent local constitutive
 tensor $\capC^{eq}$, the mutual energy simplifies in the local contribution
\beq\label{mutualmutualSGE}
\ds\mathcal{W}^{SGE}_{eq}(\overline{\bu}^\alpha,D\overline{\bu}^\alpha;\overline{\bu}^{\beta^{\diamond*}},D\overline{\bu}^{\beta^{\diamond*}})=
\alpha_{ij}\capC_{ijhk}^{eq}\int_{\Omega_{eq}}u^{\beta^{\diamond*}}_{h,k}(\bx).
\eeq
Through two applications of the divergence theorem
and using the geometrical property \textbf{GP1} of the SGE,
the mutual energy (\ref{mutualmutualSGE}) is null  and then
proposition (\ref{energy1sge}) follows. $\Box$

\subsection{Lemma 3: $\bbeta$ term in the strain energy $\mathcal{W}^{C}_{RVE}$
at first-order in $f$}\label{lemma3subsection}

\paragraph{Statement.} When a quadratic displacement $\overline{\bu}^{\beta^{\diamond *}}$,
eqn (\ref{quadraticdisplacementasterisk}) with $\balpha=\b0$,
is applied on the RVE boundary,
the strain energy at first-order in the concentration $f$ is given by eqn (\ref{jonsi}).

\paragraph{Proof.}
The strain energy $\mathcal{W}^C_{RVE}(\overline{\bu}^{\beta^{\diamond *}})$ stored in the RVE,
when a quadratic displacement field $\overline{\bu}^{\beta^{\diamond *}}$
(\ref{quadraticdisplacementasterisk}) is applied on its boundary $\partial\Omega_{RVE}$,
is bounded by (Gurtin, 1972)
\beq
\lb{boundsdispl1}
\int_{\partial\Omega_{RVE}} \sigma^{SA}_{ij} n_i \overline{u}^{\beta^{\diamond *}}_j
- \mathcal{U}_{RVE}^C (\bsigma^{SA})\leq \mathcal{W}_{RVE}^C (\overline{\bu}^{\beta^{\diamond *}})\leq \mathcal{W}_{RVE}^C (\bvarepsilon^{KA}),
\eeq
where $\bvarepsilon^{KA}$ is a kinematically admissible (satisfying the kinematic compatibility relation (\ref{kinematical})$_1$ and
the imposed displacement boundary conditions) strain field,
$\bsigma^{SA}$ is a statically admissible (satisfying the equilibrium condition, eqn (\ref{indequilibrium}) with $\btau=\b0$) stress field,
while
$\mathcal{U}_{RVE}^C (\bsigma^{SA})$ and $\mathcal{W}_{RVE}^C (\bvarepsilon^{KA})$ are respectively the following stress and strain energies
\beq
\lb{funzionirve}
\begin{array}{lll}
\ds\mathcal{U}_{RVE}^C (\bsigma^{SA})=\frac{1}{2}\int_{\Omega_R}\sigma^{SA}_{ij}(\bx) \capC^{-1}_{ijhk}(\bx)\sigma^{SA}_{hk}(\bx),\\[6mm]
\ds\mathcal{W}_{RVE}^C (\bvarepsilon^{KA})=\frac{1}{2}\int_{\Omega_R}\varepsilon^{KA}_{ij}(\bx) \capC_{ijhk}(\bx)\varepsilon^{KA}_{hk}(\bx).
\end{array}
\eeq

Considering the kinematically admissible strain field
\beq\lb{epsilonKA}
\varepsilon^{KA}_{ij}=(\beta^{\diamond *}_{ijk}+\beta^{\diamond *}_{jik})x_k,
\eeq
and assuming the geometrical properties \textbf{GP2} and \textbf{GP3},
an estimate for the {\it upper bound} in eqn (\ref{boundsdispl1}) is
the strain energy $\mathcal{W}_{RVE}^C (\bvarepsilon^{KA})$ given by eqn (\ref{strainenergyrve})$_1$ (Appendix \ref{energieboundsrve}), so that
\beq\lb{UB}
\mathcal{W}_{RVE}^{C}(\overline{\bu}^{\beta^{\diamond *}})\leq
2\rho^2 \Omega\capC^{(1)}_{ijhk}\beta^{\diamond *}_{ijl}\beta^{\diamond *}_{hkl}+o(f).
\eeq

Considering now the statically admissible stress field
\beq
\lb{stressSA}
\sigma^{SA}_{ij}=2\capC^*_{ijhk}\beta^{\diamond *}_{hkl}x_l,
\eeq
where $\capC^*$ is a first-order perturbation in $f$ to the material matrix $\capC^{(1)}$, eqn (\ref{Cauxiliary0}),
and
assuming the geometrical property \textbf{GP2}, the stress energy $\mathcal{U}^C_{RVE} (\bsigma^{SA})$
is given by eqn (\ref{strainenergyrve})$_2$ (Appendix \ref{energieboundsrve}). Moreover,
since the application of the divergence theorem yields
\beq
\lb{boundsdispl01}
\int_{\partial\Omega_R} \sigma^{SA}_{ij} n_i \overline{u}^{\beta^{\diamond *}}_j
=4\rho^2 \Omega\left(\capC^{(1)}_{ijhk}+f \hat{\capC}_{ijhk}\right)\beta^{\diamond *}_{ijl}\beta^{\diamond *}_{hkl},
\eeq
an estimate is obtained for the {\it lower bound} in eqn (\ref{boundsdispl1})
as
\beq
\lb{LB}
\mathcal{W}_{RVE}^{C}(\overline{\bu}^{\beta^{\diamond *}})\geq
2\rho^2 \Omega\capC^{(1)}_{ijhk}\beta^{\diamond *}_{ijl}\beta^{\diamond *}_{hkl}+o(f),
\eeq
which, together with the upper bound (\ref{UB}), leads to eqn (\ref{jonsi}). $\Box$

\subsection{Lemma 4: $\bbeta$ term in the strain energy $\mathcal{W}^{SGE}_{eq}$
at first-order in $f$.}\label{lemma4subsection}

\paragraph{Statement.} When a quadratic displacement $\overline{\bu}^{\beta^{\diamond *}}$,
eqn (\ref{quadraticdisplacementasterisk}) with $\balpha=\b0$,
and the normal component of its gradient $D\overline{\bu}^{\beta^{\diamond *}}$
are imposed on the boundary of the homogeneous SGE equivalent material,
 the strain energy at first-order in the concentration $f$ is given by eqn (\ref{alex}).

\paragraph{Proof.} The strain energy $\mathcal{W}^{SGE}_{eq}(\overline{\bu}^{\beta^{\diamond *}},D\overline{\bu}^{\beta^{\diamond *}})$
stored in the SGE, when a quadratic displacement field $\overline{\bu}^{\beta^{\diamond *}}$
(\ref{quadraticdisplacementasterisk}) and the normal component of its gradient $D\overline{\bu}^{\beta^{\diamond *}}$  are imposed
 on its boundary $\partial\Omega_{eq}$, is bounded as (Appendix \ref{boundsgurtinforSGM})
\beq\begin{array}{ccc}
\lb{boundsdispl}
\ds\int_{\partial\Omega_{eq}} \left(t^{SA}_i \overline{u}^{\beta^{\diamond *}}_i+
T^{SA}_i D\overline{u}^{\beta^{\diamond *}}_i\right) +
\int_{\Gamma_{eq}}\Theta_i^{SA}  \overline{u}^{\beta^{\diamond *}}_i  - \mathcal{U}^{SGE}_{eq} (\bsigma^{SA},\btau^{SA})
\leq \\[6 mm]
\ds\leq\mathcal{W}^{SGE}_{eq} (\overline{\bu}^{\beta^{\diamond *}},D\overline{\bu}^{\beta^{\diamond *}})
\leq \mathcal{W}^{SGE}_{eq} (\bvarepsilon^{KA},\bchi^{KA}),
\end{array}
\eeq
with
\beq\lb{boundaryconditions1}
\left\{
\begin{split}
t^{SA}_k&=n_j\sigma^{SA}_{jk}-n_i n_j D\tau^{SA}_{ijk}-2 n_j D_i \tau^{SA}_{ijk}+
\left(n_i n_j D_l n_l-D_j n_i\right)\tau^{SA}_{ijk},\\[3 mm]
T^{SA}_{k}&=n_i n_j \tau^{SA}_{ijk},
\end{split}
\right. \qquad \mbox{on } \partial\Omega_{eq},
\eeq
and
\beq
\lb{boundaryconditions2b}
\Theta^{SA}_k=\salto{0.5}{\, e_{mlj} n_i s_m n_l \tau_{ijk}^{SA} \,}\,\,,
\qquad ~~~ \mbox{on}\, \Gamma_{eq},
\eeq
where $\bvarepsilon^{KA}$ and $\bchi^{KA}$ are kinematically admissible
strain and curvature fields
(satisfying the kinematic compatibility relation (\ref{kinematical})
and the imposed displacement boundary conditions),
$\bsigma^{SA}$ and $\btau^{SA}$ are statically admissible
stress and double-stress fields  (satisfying the equilibrium equation (\ref{indequilibrium})), while
$\mathcal{U}^{SGE}_{eq} (\bsigma^{SA},\btau^{SA})$ and
$\mathcal{W}^{SGE}_{eq} (\bvarepsilon^{KA},\bchi^{KA})$ are respectively the stress and the strain energies given by
\beq
\lb{funzionisge}
\begin{array}{lll}
\ds\mathcal{U}^{SGE}_{eq} (\bsigma^{SA},\btau^{SA})=
\frac{1}{2}\int_{\Omega_{eq}}\sigma^{SA}_{ij}(\bx) \capC^{eq^{-1}}_{ijhk}\sigma^{SA}_{hk}(\bx)+\frac{1}{2}
\int_{\Omega_{eq}}\tau^{SA}_{ijh}(\bx) \capA^{{eq}^{-1}}_{ijhklm}\tau^{SA}_{klm}(\bx),\\[5mm]
\ds\mathcal{W}^{SGE}_{eq} (\bvarepsilon^{KA},\bchi^{KA})=
\frac{1}{2}\int_{\Omega_{eq}}\varepsilon^{KA}_{ij}(\bx) \capC^{eq}_{ijhk}\varepsilon^{KA}_{hk}(\bx)+\frac{1}{2}
\int_{\Omega_{eq}}\chi^{KA}_{ijh}(\bx) \capA^{eq}_{ijhklm}\chi^{KA}_{klm}(\bx).
\end{array}
\eeq

Considering the kinematically admissible strain $\bvarepsilon^{KA}$ (\ref{epsilonKA})
and curvature field
\beq\lb{chiKASGE}
\chi^{KA}_{ijk}=2\beta^{\diamond *}_{kij},
\eeq
and assuming geometrical property \textbf{GP2},
an estimate for the {\it upper bound} in eqn (\ref{boundsdispl})
is the strain energy
$\mathcal{W}^{SGE}_{eq} (\bvarepsilon^{KA},\bchi^{KA})$
given by eqn (\ref{strainenergysge})$_1$ (Appendix \ref{energieboundssge}) as
\beq
\lb{UBSGE}
\mathcal{W}^{SGE}_{eq} (\overline{\bu}^{\beta^{\diamond *}},D\overline{\bu}^{\beta^{\diamond *}})\leq
 2\Omega\beta^{\diamond *}_{ijl}\beta^{\diamond *}_{hkm}
\left(\rho^2\capC^{eq}_{ijhk}\delta_{lm}+\capA^{eq}_{jlikmh}\right).
\eeq
Considering the  statically admissible stress  $\bsigma^{SA}$ (\ref{stressSA}) and
double-stress field
\beq\lb{tauSASGE}
\tau^{SA}_{jli}=2\capA^{eq}_{jlikmh}\beta^{\diamond *}_{hkm},
\eeq
where $\capC^*$ is a first-order perturbation in $f$ to the material matrix $\capC^{eq}$, eqn (\ref{CauxiliarySGE}) and
assuming the geometrical property \textbf{GP2}, the stress energy $\mathcal{U}^{SGE}_{eq} (\bsigma^{SA},\btau^{SA})$
is given by eqn (\ref{ultimosforzo}) (Appendix \ref{energieboundssge}). Moreover,
since the application of the divergence theorem yields
\beq
\lb{boundsdisplSGE}
\begin{array}{lll}
&\ds\int_{\partial\Omega_{eq}} \left(t^{SA}_i \overline{u}^{\beta^{\diamond *}}_i+
T^{SA}_i D\overline{u}^{\beta^{\diamond *}}_i\right) +
\int_{\Gamma_{eq}}\Theta_i^{SA}  \overline{u}^{\beta^{\diamond *}}_i=
4\rho^2\Omega\left[\capC^{eq}_{ijhk}+
f \left(\hat{\capC}_{ijhk}-\tilde{\capC}_{ijhk}\right)\right]\beta^{\diamond *}_{ijn}\beta^{\diamond *}_{hkn},
\end{array}
\eeq
an estimate is obtained for the {\it lower bound} in eqn (\ref{boundsdispl}) as
\beq
\lb{LBSGE}
\mathcal{W}^{SGE}_{eq} (\overline{\bu}^{\beta^{\diamond *}},D\overline{\bu}^{\beta^{\diamond *}})\geq
 2\Omega\beta^{\diamond *}_{ijl}\beta^{\diamond *}_{hkm}
\left(\rho^2\capC^{eq}_{ijhk}\delta_{lm}+\capA^{eq}_{jlikmh}\right)+o(f),
\eeq
which, together with the upper bound (\ref{UBSGE}), leads to eqn (\ref{alex}). $\Box$

\section{Elastic energies based on the kinematically admissible
displacement field $\bu^{\beta^{\diamond *}}$ (\ref{quadraticdisplacementinbodyasterisk})}

\setcounter{equation}{0}
\renewcommand{\theequation}{{B}.\arabic{equation}}

In this Appendix it is assumed $\balpha=\b0$.
The field $\bu^{\beta^{\diamond *}}$, eqn (\ref{quadraticdisplacementinbodyasterisk}), is
a kinematically admissible displacement
for both boundary conditions $\overline{\bu}^{\beta^{\diamond *}}$, eqn (\ref{bcrve}),
 and ($\overline{\bu}^{\beta^{\diamond *}}$, $D\overline{\bu}^{\beta^{\diamond *}}$), eqn (\ref{bcsge}),
applied on the boundary of the RVE and the SGE, respectively.
The related strain and stress energies in the RVE and in the SGE are obtained below.
\begin{itemize}
\item In Section B.1 the strain energies are computed with the kinematically admissible deformation $\bvarepsilon^{KA}$, eqn (\ref{epsilonKA}),
and curvature $\bchi^{KA}$, eqn  (\ref{chiKASGE}), originated by the kinematically
admissible displacement $\bu^{\beta^{\diamond *}}$, eqn (\ref{quadraticdisplacementinbodyasterisk});
\item In Section B.2 the stress energies are computed with the statically admissible stress $\bsigma^{SA}$, eqn (\ref{stressSA}),
and double-stress $\btau^{SA}$, eqn  (\ref{tauSASGE}), originated by the above mentioned kinematically
admissible fields $\bvarepsilon^{KA}$ and$\bchi^{KA}$
within a homogeneous material with constitutive tensors $\capC^*$ and $\capA^{eq}$.
\end{itemize}

\subsection{Strain and stress energies in the RVE}\label{energieboundsrve}

The kinematically admissible deformation $\bvarepsilon^{KA}$, eqn (\ref{epsilonKA}),
and the statically admissible stress $\bsigma^{SA}$, eqn (\ref{stressSA}), provide
the strain and stress energies (\ref{funzionirve}) in the RVE
\beq\begin{array}{lll}
\ds\mathcal{W}^{C}_{RVE}(\bvarepsilon^{KA})=&\ds\int_\Omega
2\capC_{ijhk}(\bx)\beta^{\diamond *}_{ijl}\beta^{\diamond *}_{hkm} x_l x_m,\\[4mm]
\ds\mathcal{U}^{C}_{RVE}(\bsigma^{SA})=&\ds\int_\Omega
2\capC^*_{ijlm}\capC^{-1}_{ijhk}(\bx)\capC^*_{hkrs}\beta^{\diamond *}_{lmn}\beta^{\diamond *}_{rst}  x_n x_t,
\end{array}
\eeq
which, introducing the definition (\ref{euler}) of the Euler tensor of inertia $\bE$, can be rewritten as
\beq\label{dai}
\begin{array}{lll}
\mathcal{W}^{C}_{RVE}(\bvarepsilon^{KA})=&
\ds2\left[\capC_{ijhk}^{(1)}E_{lm}(\Omega_{1}^C)+
\capC_{ijhk}^{(2)}E_{lm}(\Omega_{2}^C)\right]\beta^{\diamond *}_{ijl}\beta^{\diamond *}_{hkm},\\[4mm]
\mathcal{U}^{C}_{RVE}(\bsigma^{SA})=&
\ds2\capC^*_{ijlm}\left\{\capC_{ijhk}^{(1)^{-1}}E_{nt}(\Omega_{1}^C)+
\capC_{ijhk}^{(2)^{-1}}E_{nt}(\Omega_{2}^C)\right\}\capC^*_{hkrs}\beta^{\diamond *}_{lmn}\beta^{\diamond *}_{rst}.
\end{array}
\eeq
Assuming the geometrical property \textbf{GP2}
and considering the identity (\ref{legamerho}), the strain and stress energies (\ref{dai}) simplify as
\beq\begin{array}{lll}
\mathcal{W}^{C}_{RVE}(\bvarepsilon^{KA})=&
\ds2\rho^2\Omega
\left\{ \capC^{(1)}_{ijhk}-f\left(\frac{\rho^{(2)}}{\rho}\right)^2
\left[\capC^{(1)}_{ijhk}-\capC^{(2)}_{ijhk}\right]\right\}\beta^{\diamond *}_{ijl}\beta^{\diamond *}_{hkl},\\[4mm]
\mathcal{U}^{C}_{RVE}(\bsigma^{SA})=&
\ds2\rho^2\Omega\capC^*_{ijlm}
\left\{\capC_{ijhk}^{(1)^{-1}}-f\left(\frac{\rho^{(2)}}{\rho}\right)^2
\left[\capC_{ijhk}^{(2)^{-1}}-\capC_{ijhk}^{(1)^{-1}}\right]\right\}
\capC^*_{hkrs}\beta^{\diamond *}_{lmn}\beta^{\diamond *}_{rsn}.
\end{array}
\eeq
Assuming the geometrical property \textbf{GP3}
\beq
\rho^{(2)}= \tilde\rho^{(2)} f^{r}+o(f),
\eeq
with $0<r\leq1$, and $\capC^*$ as a first-order perturbation
in $f$ to the material matrix $\capC^{(1)}$, eqn (\ref{Cauxiliary0}), the strain and the stress energies
are given in the dilute case ($f\ll1$) by
\beq\label{strainenergyrve}
\begin{array}{lll}
\mathcal{W}^{C}_{RVE}(\bvarepsilon^{KA})=&
\ds2\rho^2\Omega
\capC^{(1)}_{ijhk}\beta^{\diamond *}_{ijl}\beta^{\diamond *}_{hkl}+o(f),\\[4mm]
\mathcal{U}^{C}_{RVE}(\bsigma^{SA})=&
\ds2\rho^2\Omega\left(\capC^{(1)}_{ijhk}+2f\hat\capC_{ijhk}\right)\beta^{\diamond *}_{ijl}\beta^{\diamond *}_{hkl}+o(f).
\end{array}
\eeq

\subsection{Strain and stress energies in the SGE}\label{energieboundssge}

The  kinematically admissible deformation and curvature fields [$\bvarepsilon^{KA}$, eqn (\ref{epsilonKA});
$\bchi^{KA}$, eqn  (\ref{chiKASGE})] together with
the statically admissible stress and double-stress fields [$\bsigma^{SA}$, eqn (\ref{stressSA});
 $\btau^{SA}$, eqn  (\ref{tauSASGE})]
provide the strain and stress energies (\ref{funzionisge}) in the SGE
\beq\begin{array}{lll}
\mathcal{W}^{SGE}_{eq}(\bvarepsilon^{KA}, \bchi^{KA})=&\ds\int_\Omega
2\left[\capC^{eq}_{ijhk} x_l x_m + \capA^{eq}_{jlikmh}\right]\beta^{\diamond *}_{ijl}\beta^{\diamond *}_{hkm},\\[4mm]
\ds\mathcal{U}^{SGE}_{eq}(\bsigma^{SA}, \btau^{SA})=&\ds\int_\Omega
2\left\{\capC^*_{ijlm}\capC^{eq^{-1}}_{ijhk}\capC^*_{hkrs}x_n x_t+
\capA^{eq}_{mnlstr}\right\}\beta^{\diamond *}_{lmn}\beta^{\diamond *}_{rst},
\end{array}
\eeq
which, introducing the definition (\ref{euler}) for the Euler tensor of inertia $\bE$, can be rewritten as
\beq\label{daidai}\begin{array}{lll}
\mathcal{W}^{SGE}_{eq}(\bvarepsilon^{KA}, \bchi^{KA})=&
\ds2\left[
\capC_{ijhk}^{eq}E_{lm}(\Omega_{eq}^{SGE})+\Omega_{eq}^{SGE}\capA^{eq}_{jlikmh}\right]\beta^{\diamond *}_{ijl}\beta^{\diamond *}_{hkm},\\[4mm]
\ds\mathcal{U}^{SGE}_{eq}(\bsigma^{SA}, \btau^{SA})=&\ds
2\left\{\capC^*_{ijlm}\capC^{eq^{-1}}_{ijhk}\capC^*_{hkrs}E_{nt}(\Omega_{eq}^{SGE})+\Omega_{eq}^{SGE}
 \capA^{eq}_{mnlstr}\right\}\beta^{\diamond *}_{lmn}\beta^{\diamond *}_{rst}.
\end{array}
\eeq
Assuming the geometrical property \textbf{GP2},
the strain and stress energies (\ref{daidai}) simplify as
\beq\label{strainenergysge}
\begin{array}{lll}
\mathcal{W}^{SGE}_{eq}(\bvarepsilon^{KA}, \bchi^{KA})=&
\ds2
\Omega\left[\rho^2\capC_{ijhk}^{eq}\delta_{lm}+\capA^{eq}_{jlikmh}
\right]
\beta^{\diamond *}_{ijl}\beta^{\diamond *}_{hkm},\\[4mm]
\ds\mathcal{U}^{SGE}_{eq}(\bsigma^{SA}, \btau^{SA})=&\ds
2\Omega\left\{\rho^2\capC^*_{ijlm}\capC^{eq^{-1}}_{ijhk}\capC^*_{hkrs}\delta_{nt}+
 \capA^{eq}_{mnlstr}\right\}\beta^{\diamond *}_{lmn}\beta^{\diamond *}_{rst}.
\end{array}
\eeq
Finally, assuming  $\capC^*$ as a first-order perturbation
in $f$ to the equivalent local tensor $\capC^{eq}$, eqn (\ref{CauxiliarySGE}),
the stress energy is given in the dilute case ($f\ll1$) by
\beq\label{ultimosforzo}\begin{array}{lll}
\ds\mathcal{U}^{SGE}_{eq}(\bsigma^{SA}, \btau^{SA})=&\ds
2\Omega\left\{\rho^2\left[\capC^{eq}_{ijhk}+2f\left(\hat\capC_{ijhk}-\tilde\capC_{ijhk}\right)\right]\delta_{nt}+
 \capA^{eq}_{mnlstr}\right\}\beta^{\diamond *}_{lmn}\beta^{\diamond *}_{rst}+o(f).
\end{array}
\eeq

\section{Energy bounds for SGE Material}\label{boundsgurtinforSGM}

\setcounter{equation}{0}
\renewcommand{\theequation}{{C}.\arabic{equation}}

\paragraph{Statement.} When boundary displacement conditions $\overline{\bu}$,
$\overline{D\bu}$ are imposed on the boundary $\partial\Omega_{eq}$ of a SGE,
the strain energy $\mathcal{W}^{SGE}_{eq} (\overline{\bu},\overline{D\bu})$
is bounded as
\beq\label{bandus}
\begin{array}{ccc}
\ds\int_{\partial\Omega_{eq}} \left(t^{SA}_i \overline{u}_i+
T^{SA}_i \overline{D u}_i\right) +
\int_{\Gamma_{eq}}\Theta_i^{SA}  \overline{u}_i  - \mathcal{U}^{SGE}_{eq} (\bsigma^{SA},\btau^{SA})
\leq \mathcal{W}^{SGE}_{eq} (\overline{\bu},\overline{D\bu})
\leq \mathcal{W}^{SGE}_{eq} (\bvarepsilon^{KA},\bchi^{KA}),
\end{array}
\eeq
where $\bvarepsilon^{KA}$ and $\bchi^{KA}$ are kinematically admissible
strain and curvature fields
(satisfying the kinematic compatibility relation (\ref{kinematical})
and the imposed displacement boundary conditions), $\bsigma^{SA}$ and $\btau^{SA}$ are statically admissible
stress and double-stress fields (satisfying the equilibrium equation (\ref{indequilibrium}))
and the other statically admissible quantities $\bt^{SA}$, $\bT^{SA}$
and $\bTheta^{SA}$ are given by eqns (\ref{boundaryconditions1}) and (\ref{boundaryconditions2b}), while
$\mathcal{U}^{SGE}_{eq} (\bsigma^{SA},\btau^{SA})$ and
$\mathcal{W}^{SGE}_{eq} (\bvarepsilon^{KA},\bchi^{KA})$ are respectively the stress and the strain energies, eqns
(\ref{funzionisge})$_1$ and (\ref{funzionisge})$_2$.

\paragraph{Proof.} Considering the displacement field $\bu^{eq}$ solution
to the displacement boundary conditions $\overline{\bu}$, $\overline{D\bu}$
and the related statical fields $\bsigma^{eq}$ and $\btau^{eq}$ in equilibrium,
through the difference fields $\Delta\bvarepsilon^{KA}$, $\Delta\bchi^{KA}$,
$\Delta\bsigma^{SA}$, $\Delta\btau^{SA}$
the kinematically and statically admissible fields can be defined as
\beq\begin{array}{lll}
\bvarepsilon^{KA}=\bvarepsilon^{eq}+\Delta\bvarepsilon^{KA},\qquad
\bchi^{KA}=\bchi^{eq}+\Delta\bchi^{KA},\\[4mm]
\bsigma^{SA}=\bsigma^{eq}+\Delta\bsigma^{SA},\qquad
\btau^{SA}=\btau^{eq}+\Delta\btau^{SA}.
\end{array}
\eeq
Using the discrepancy fields $\Delta\bvarepsilon^{KA}$ and $\Delta\bchi^{KA}$
 the term representing the {\it upper bound} in eqn (\ref{bandus}) can be rewritten as
\beq\label{proofboundup}
\begin{array}{lll}\ds
\mathcal{W}^{SGE}_{eq} (\bvarepsilon^{KA},\bchi^{KA})=&
\mathcal{W}^{SGE}_{eq} (\overline{\bu},\overline{D\bu})+
\mathcal{W}^{SGE}_{eq} (\Delta\bvarepsilon^{KA},\Delta\bchi^{KA})\\[4 mm]
&+\ds\int_{\Omega_{eq}}\left(\capC_{ijhk}\varepsilon_{ij}^{eq}\Delta\varepsilon_{hk}^{KA}
+\capA_{ijklmn}\chi_{ijk}^{eq}\Delta\chi_{lmn}^{KA}\right),
\end{array}
\eeq
which provides a proof to the upper bound, since the strain energy is positive definite and the
third term in the RHS of eqn (\ref{proofboundup}) is null by the principle of virtual
work (\ref{pvw}) with $\Delta\bu=\Delta D\bu=\b0$ on the boundary.

Using the discrepancy fields $\Delta\bsigma^{KA}$ and $\Delta\btau^{KA}$
 the term representing the {\it lower bound} in eqn (\ref{bandus}) can be rewritten as
\beq\begin{array}{ccc}
\ds\int_{\partial\Omega_{eq}} \left(t^{SA}_i \overline{u}_i+
T^{SA}_i \overline{D u}_i\right) +
\int_{\Gamma_{eq}}\Theta_i^{SA}  \overline{u}_i  - \mathcal{U}^{SGE}_{eq} (\bsigma^{SA},\btau^{SA})
=\mathcal{W}^{SGE}_{eq} (\overline{\bu},\overline{D\bu})
-\mathcal{U}^{SGE}_{eq} (\Delta\bsigma^{SA},\Delta\btau^{SA})
\end{array}
\eeq
which provides a proof to the lower bound, since the strain energy is positive definite. $\Box$


\begin{thebibliography}{}



\bibitem{ache}  Achenbach, J. D., and Herrmann, G. (1968) Dispersion of Free Harmonic
Waves in Fibre-Reinforced Composites, \emph{AIAA J.}, 6, 1832-–1836.

\bibitem{anderson}  Anderson, W. B. and Lakes, R. S. (1994)
Size effects due to Cosserat elasticity and surface damage in closed-cell polymethacrylimide foam,
\emph{J. Mat. Sci.}, 29, 6413--6419.

\bibitem{au}Auffray, N., Bouchet, R.  and Brechet, Y.  (2010) Strain gradient elastic homogenization
of bidimensional cellular media \IJSS 47, 1698--1710.

\bibitem{beran}  Beran, M. J., and McCoy, J. J. (1970) Mean Field Variations in a Statistical
Sample of Heterogeneous Linearly Elastic Solids, \IJSS, \textbf{6}, 1035-–1054.


\bibitem{bacca}  Bacca, M., Bigoni, D., Dal Corso, F. and Veber, D. (2012)
Mindlin second-gradient elastic properties from dilute two-phase Cauchy-elastic composites.
Part II: Higher-order constitutive properties and application cases. \IJSS {\it Submitted}.



\bibitem{Bigoni} Bigoni, D., and Drugan, W.J. (2007) Analytical derivation of Cosserat moduli via homogenization of heterogeneous
elastic materials. \JAM, 74, 741--753.

\bibitem{bout} Boutin, C. (1996) Microstructural effects in elastic composites. \IJSS 33, 1023-1051.

\bibitem{bouyge}  Bouyge, F., Jasiuk, I., and Ostoja-Starzewski, M. (2001) A Micromechanically
Based Couple-Stress Model of an Elastic Two-Phase Composite, \IJSS 38, 1721-–1735.

\bibitem{buechner}  Buechner, P.M., and Lakes, R.S. (2003)
Size effects in the elasticity and viscoelasticity of bone,
\emph{Biomech. Model. Mech.}, 1, 295--301.


\bibitem{cosserat} Cosserat, E., and Cosserat, F. (1909) {\it Sur la th\'{e}orie des corps d\'{e}formables}, Herman, Paris.


\bibitem{dalcor} Dal Corso, F. and Deseri. L. (2013)
Residual stresses in random elastic composites: nonlocal micromechanics-based models and first estimates of the representative volume element size.
Meccanica, In Press.


\bibitem{dalcorso} Dal Corso, F. and Willis, J.R. (2011) Stability of strain gradient plastic materials.
\JMPS, \textbf{59}, 1251--1267.

\bibitem{Eshelby} Eshelby, J.D., (1957) The Determination of the Elastic Field of an Ellipsoidal Inclusion and Related Problems.
\PRSA, 241, 376-–396.


\bibitem{forest} Forest, S. (1998) Mechanics of Generalized Continua: Construction by Homogenization, \emph{J. Phys. IV}, 8, 39-–48.


\bibitem{for} Forest, S. and Trinh, D.K. (2011) Generalized continua and non-homogeneous boundary
conditions in homogenisation methods. \ZAMM 91, 90-109.

\bibitem{gauthier} Gauthier, R.D. (1982) {\it Experimental Investigation on Micropolar Media},
Mechanics of Micropolar Media, O. Brulin and R. K. T. Hsieh, eds., CISM
Lecture Notes, World Scientific, Singapore, 395-–463.


\bibitem{gurtin} Gurtin, M.E. (1972) The linear theory of Elasticity. In Flugge, S., ed., Encyclopedia
of Physics VIa/2. Berlin, Springer. 1-295.


\bibitem{Hashin} Hashin, Z. (1959) {\it The Moduli of an Elastic Solid Containing Spherical Particles of Another Elastic Material}.
Non-Homogeneity in Elasticity and Plasticity, W. Olszak, ed., Pergamon, New York, pp. 463-–478.

\bibitem{koiter} Koiter, W.T. (1964) Couple-Stresses in the Theory of Elasticity, Parts I and
II. {\it Proc. K. Ned. Akad. Wet.}, Ser. B: Phys. Sci., 67, 17-–44.

\bibitem{lakes} Lakes, R.S. (1986) Experimental Microelasticity of Two Porous Solids, \IJSS 22, 55-–63.

\bibitem{li} Li, J. (2011) A micromechanics-based strain gradient damage model for fracture prediction
of brittle materials. Part I: Homogenization methodology and constitutive relations. \IJSS 48, 3336-3345.



\bibitem{Mindlin} Mindlin, R.D. (1964) Micro-structure in linear elasticity. \emph{Archs ration. Mech. Analysis} \textbf{16}, 51--78.

\bibitem{Mindlin-Eshel} Mindlin, R.D. and Eshel, N.N. (1968) On First Strain-Gradient Theories in Linear Elasticity. \IJSS \textbf{4}, 109.

\bibitem{Ostoja} Ostoja-Starzewski, M., Boccara, S., and Jasiuk, I. (1999) Couple-Stress
Moduli and Characteristic Length of Composite Materials, \MRC 26, 387–397.

\bibitem{pideri} Pideri, C., and Seppecher, P. (1997) A second gradient material resulting from the homogenization
of an heterogeneous linear elastic medium. \CMT 9, 241-–257.

\bibitem{wang}  Wang, X.L., and Stronge, W.J. (1999), Micropolar Theory for Two-
Dimensional Stresses in Elastic Honeycomb, \PRSA, 445, 2091-–2116.





\end{thebibliography}
\end{document}